\documentclass[preprint]{aastex}
\usepackage{graphicx}
\usepackage{natbib}
\usepackage{latexsym}
\usepackage{amssymb}
\usepackage{longtable}
\def\msun{\ifmmode {\rm\,M_\odot}\else ${\rm\,M_\odot}$\fi}
\def\Msun{\ifmmode {\rm\,\it{M_\odot}}\else ${\rm\,M_\odot}$\fi}
\def\lsun{\ifmmode {\rm\,L_\odot}\else ${\rm\,L_\odot}$\fi}
\def\Lsun{\ifmmode {\rm\,\it{L_\odot}}\else ${\rm\,L_\odot}$\fi}
\def\rsun{\ifmmode {\rm\,R_\odot}\else ${\rm\,R_\odot}$\fi}
\def\Rsun{\ifmmode {\rm\,\it{R_\odot}}\else ${\rm\,R_\odot}$\fi}
\def\Tsun{\ifmmode {\rm\,T_\odot}\else ${\rm\,T_\odot}$\fi}
\def\arcsec{\ifmmode {^{\prime\prime}}\else $^{\prime\prime}$\fi}
\def\asec{\ifmmode {^{\prime\prime}}\else $^{\prime\prime}$\fi}
\def\arcmin{\ifmmode {^{\prime}}\else $^{\prime}$\fi}
\def\amin{\ifmmode {^{\prime}}\else $^{\prime}$\fi}
\def\simlt{\mathrel{\spose{\lower 3pt\hbox{$\mathchar"218$}}
     \raise 2.0pt\hbox{$\mathchar"13C$}}}
\def\simgt{\mathrel{\spose{\lower 3pt\hbox{$\mathchar"218$}}
\     \raise 2.0pt\hbox{$\mathchar"13E$}}}

\begin{document}

\title{Testing Disk--Locking in NGC 2264}
\author{P. Wilson Cauley, Christopher M. Johns--Krull\altaffilmark{1,2}}
\email{pwc1@rice.edu, cmj@rice.edu}
\affil{Rice University}
\affil{Department of Physics and Astronomy, 6100 Main St., MS 108, Houston, TX 77005}
\author{Catrina M. Hamilton\altaffilmark{1}}
\affil{Dickinson College}
\affil{Department of Physics and Astronomy, P.O. Box 1773, Dickinson College, Carlisle, PA 17013}
\email{hamiltoc@dickinson.edu}
\and
\author{Kelly Lockhart}
\affil{University of Hawaii Manoa}
\affil{Institute for Astronomy, 2680 Woodlawn Drive, University of Hawaii, Honolulu, HI 96822}
\email{kel@ifa.hawaii.edu}

\altaffiltext{1}{Visiting Astronomer, McDonald Observatory, operated
by The University of Texas at Austin, Austin, Texas, USA}

\altaffiltext{2}{Visiting Astronomer, Kitt Peak National Observatory, National Optical Astronomy Observatory, which is operated by the Association of Universities for Research in Astronomy (AURA) under cooperative agreement with the National Science Foundation.}

\begin{abstract}
We test analytic predictions from different models of magnetospheric accretion, which invoke disk--locking, using stellar and accretion parameters derived from models of low resolution optical spectra of 36 T Tauri stars (TTSs) in NGC 2264 (age$\sim$3 Myrs).  Little evidence is found for models that assume purely dipolar field geometries; however, strong support is found in the data for a modified version of the X--wind model \citep{shu94} which allows for non--dipolar field geometries.  The trapped flux concept in the X--wind model is key to making the analytic predictions which appear supported in the data.  By extension, our analysis provides support for the outflows predicted by the X--wind as these also originate in the trapped flux region.  In addition, we find no support in the data for accretion powered stellar winds from young stars.  By comparing the analysis presented here of NGC 2264 with a similar analysis of stars in Taurus (age$\sim$1--2 Myr), we find evidence that the equilibrium interaction between the magnetic field and accretion disk in TTS systems evolves as the stars grow older, perhaps as the result of evolution of the stellar magnetic field geometry.  We compare the accretion rates we derive with accretion rates based on U--band excess, finding good agreement.  In addition, we use our accretion parameters to determine the relationship between accretion and H$\beta$ luminosity, again finding good agreement with previously published results; however, we also find that care must be used when applying this relationship due to strong chromospheric emission in young stars which can lead to erroneous results in some cases.                        
\end{abstract}

\newpage

\section{INTRODUCTION}

T Tauri stars (TTSs), first classified and named by \citet{Joy45}, are low mass, pre--main sequence (PMS) objects that are roughly grouped into two classifications: 1. classical T Tauri stars (CTTSs), which show evidence of a circumstellar disk and mass accretion onto the central star in the form of excess emission in the X--ray, UV, optical, and infrared \citep{bert89,argiroffi07}, and 2. weak--line T Tauri stars (WTTSs), or naked T Tauri stars (NTTSs), which are also PMS stars but do not show evidence for significant mass accretion and do not have inner disks \citep{walt87}.  WTTSs are thought to be the evolutionary product of CTTSs that have ceased accreting material from their disks and are continuing their gravitational contraction to the main sequence, though there is significant overlap in HR diagrams.  Understanding how CTTSs interact with and ultimately disperse their disks is vital to our knowledge of how these systems evolve, supposedly into the WTTS phase, and eventually become sun--like stars, some of which are likely surrounded by planetary systems similar to our own.  The goal of this study is to test predictions of magnetospheric accretion theories (discussed below) in the $\sim$3 Myr old open cluster NGC 2264.  The slightly greater age of NGC 2264 compared to well--studied regions like Orion and the Taurus--Auriga star forming region, both with an age $\sim$1--2 Myr, allows an exploration of magnetospheric accretion in an older population of TTSs which permits an investigation of how magnetoshperic accretion may evolve over time.      

It is now well established that the excess emission observed in the spectra of CTTSs is due to the accretion of gas from a circumstellar disk \citep[e.g.][]{bert88,phart91}.  The original proposition for how the excess emission forms was put forth by \citet{lbp74} who imagined the interaction between star and disk occuring in a boundary layer: the fast--rotating disk dissipates its energy in a narrow region which is in contact with the surface of the more slowly rotating central star, resulting in the excess emission observed at blue and UV wavelengths.  While the boundary layer paradigm is able to explain the hot continuum emission \citep[e.g.][]{bert88,bb89}, it fails to account for the large velocity--shifts in the line profiles of CTTSs and the large equivalent widths of H$\alpha$ emission that are commonly observed \citep{hart98}.  Further evidence against the boundary layer model includes observed inner--disk holes around some CTTSs \citep{meyer97}, which nevertheless shows signs of accretion \citep{hart98}.  Chromospheric models explaining the observed excesses were proposed by \cite{cram79} and \cite{calvet84}, and these were successful in reproducing some of the spectral chracteristics of CTTSs.  However, like the boundary layer paradigm, these models fail to account for velocity shifts and widths in spectral lines and could not reproduce near--IR (disk) excesses, as well as becoming increasingly unfeasible when used to describe highly veiled spectra \citep{calvet84}.  Magnetospheric accretion, in which gas from the disk is loaded onto stellar magnetic field lines and impacts the surface of the star at near free--fall velocities, provides the best explanation for the observed properties of CTTSs \citep{bouvier07}.    

Support for magnetospheric accretion as the dominant accretion process on CTTSs is strong.  Magnetospheric accretion onto astrophysical objects was first investigated by \cite{gl77} who modeled pulsating neutron stars accreting mass from a binary companion.  \cite{us84} first suggested a similar accretion mechanism for TTSs.  The observational link between young stellar objects (YSOs) and the role magnetic fields play in their evolution was first established based on images of strong, collimated bipolar outflows and more highly collimated jets, and the energy required to power them \citep[see][]{apenmundt89}.  

The first investigations of mass loss from TTSs, based on observations of energetic winds, were performed by \cite{Varsavsky}, \cite{herbig61}, and \cite{kuhi64} by assuming that the flows were driven from the surface of the star.  Some of the first evidence for the \textit{circumstellar} origin of outflows in CTTSs, and as a result the conclusion that these outflows might be intrinsically linked to mass accretion onto the star, was discovered by \cite{edwards89}.  They suggested that the winds and outflows were direct results of accretion and thus could not be purely stellar in origin.  We now know that bipolar outflows are ubiquitous phenemona in YSOs \citep{bally07}.  These outflows require the existence of strong, hourglass--shaped magnetic field lines to transport and collimate material from the disk or star into the observed bipolar flows \citep{pudritz83,pudritz86,shu88,cam90}; winds driven purely by thermal or radiation pressure cannot account for the large outflow velocities that are observed \citep{lada85}.  Magnetic field lines can also act as a collimating agent, bounding the outflows at large distances from the central star \citep[e.g.][]{shu88}.  Magnetospheric accretion theories can also account for the observed variability of the UV excess \citep{bert88,alencar01}: accretion columns terminating at the stellar surface co--rotate with the star, thus producing variability on the same timescale as the stellar rotation period.  Velocity shifts of emission and absorption lines are naturally explained by the $\sim$250 km/s velocities of the infalling material \citep[e.g.][]{calvet98}.

One of the most important questions concerning the evolution of TTSs is the problem of how these objects are able to shed angular momentum and rotate with velocities well below break--up \citep[e.g.][]{vogel81,herbst02,lamm04,mak04}.  The magnetospheric accretion model provides a potential answer to this question due to the role of the stellar magnetic field in transferring angular momentum from the star to the disk: if the magnetic field couples with a sufficiently ionized disk and acts as a braking torque on the star, it will rotate more slowly than if no braking mechanism were present \citep{kon91,shu94}.  This interaction essentially locks the star to the disk (i.e. \textit{disk--locking}) and prevents the star from rotating at break--up velocity.  

This scenario can account for many of the observed rotation rates of PMS stars \citep[][hereafter L04]{edwards93,kearns98,lamm04}.  Strong evidence for disk--locked stellar rotation has been found in the Orion Nebula Cluster (ONC) \citep{choi96,herbst02} and NGC 2264 \citep[][hereafter L05]{lamm05}, though contradictory results have been found by the studies of \cite{stass99} and \cite{rebull01} for the ONC, and \cite{mak04} for NGC 2264.  Evidence supporting magnetic disk--locking manifests itself in the form of a bimodal period distribution \citep[e.g.][]{attridge92} and the detection of disk signatures \citep{edwards93,rebull06}, and hence circumstellar disks, around long--period stars.  \cite{herbst02}, based on the work of \cite{choi96}, point out that the shorter period peak in the bimodal distribution is simply a binning artifact (i.e. the rotational evolution of stars that have ceased to be regulated by their disks should span a range of shorter periods) while the longer period peak ($\sim$8 days) is a direct result of magnetic disk--locking \citep{attridge92,herbst01,herbst02}.  To date, studies by \cite{rebull06}, \cite{dahm2011}, and \cite{cieza07} provide the strongest evidence for correlations between slowly rotating stars and the presence of a circumstellar disk, though the \cite{dahm2011} study finds high confidence levels ($>$99\%) for only the M--dwarfs in their sample.  The work of \cite{herbst02} also shows strong evidence that longer period stars have a higher incidence of circumstellar disks.  These correlations support the hypothesis of star--disk interactions being the main culprit in removing angular momentum from TTS systems.\defcitealias{lamm05}{L05}\defcitealias{lamm04}{L04} 

\citetalias{lamm05} also report a bimodal period distribution for a large sample of stars in NGC 2264, similar to what has been reported for the ONC.  However, \cite{mak04} find no evidence for a bimodal period distribution in NGC 2264.  In fact, \cite{mak04} find no significant difference between the period distributions of the Orion region and NGC 2264, which differ in age by a factor of $\sim$2 \citepalias{lamm05}.  This result directly conflicts with the results of \citetalias{lamm05} who find that the period distribution of angular momentum conservation for NGC 2264 is consistent with stellar contraction from the period distribution of the ONC based on fully convective PMS models.  \citetalias{lamm05} explain that the conclusions of \cite{mak04} are a result of the latter study comparing the period distribution of NGC 2264 with a larger, inhomogeneous region in Orion.  \citetalias{lamm05}, on the other hand, only compare the period distribution of NGC 2264 with the younger homogeneous region of the ONC.  Studies of both clusters find that the rotation period distributions for higher mass stars ($M_*$ $>$ 0.25 M$_\odot$) and lower mass stars ($M_*$ $<$ 0.25 M$_\odot$) peak at different locations, with the lower mass stars peaking at a shorter period than those with greater mass \citep{herbst01,herbst02,mak04,lamm05}.

Despite some conflicting results, observational evidence seems to point to disk--star interactions as being the primary candidate for explaining TTS rotation rates.  Further support for magnetic disk--locking can be found in the measurements of magnetic fields on TTSs.  Though the sample size of measured fields is relatively small, \cite{JK99}, \cite{guenther99}, \cite{JK07}, and \cite{yangJK11} measure relatively uniform values of $\sim$1-3 kG for surface magnetic fields on TTSs.  The relative constancy of their magnetic fields, combined with the ubiquitous presence of circumstellar disks, points to interactions between the stellar magnetic field and disks as being a prime candidate for angular momentum regulation in CTTSs.  However, \cite{JK07} points out that the magnetic fields of TTSs may not be strong enough to enforce disk--locking if a dipolar field geometry is not assumed at the stellar surface.  On the other hand, \cite{JK02}, hereafter JG02, investigated correlations predicted by several different theories of magnetospheric accretion (see below) and found support for models that assume a \textit{non}--dipolar surface field geometry and only weak support for models assuming purely dipolar fields.\defcitealias{JK02}{JG02}\defcitealias{OS95}{OS95}

In this study, we extend the analysis performed by \citetalias{JK02} to a larger sample of PMS stars in NGC 2264.  \citetalias{JK02} examined several analytic relationships predicted by four different magnetospheric accretion theories, specifically those of \cite{kon91}, \cite{camcamp93}, \cite{shu94}, and a modified version of the \cite{OS95} model (hereafter OS95), using multiple sets of observations of CTTSs in the Taurus--Auriga molecular cloud complex.  The theories and their predictions will be discussed in Section 2.  Modeling the accretion columns producing the excess emission in CTTSs as hot slabs of hydrogen \citep[e.g.][]{valenti93} enables us to derive reliable estimates of stellar and accretion parameters for our sample in NGC 2264.  We then test the same correlations predicted by the aforementioned theories for this set of stars.  NGC 2264 is an ideal candidate for this study due to the availability of rotation periods and spectral type determinations for a large number of stars.  As mentioned above, NGC 2264 is slightly older than the Taurus--Auriga region and the ONC (3 Myrs vs. 1-2 Myrs; \citetalias{lamm05}) and so should provide a good testing ground for the models using a more evolved PMS stellar population.  Exploring young clusters at different ages allows us to see when the equilibrium magnetspheric accretion relationships and disk--locking break down.  Section 3 describes our observations and reductions and the classification of the stellar sample into accreting and negligibly accreting stars.  In Section 4 we discuss our models and how they are used to derive stellar and accretion parameters (see Appendix A for a detailed description of our fitting procedure and model assumptions).  Section 5 includes the application of the accretion theory predictions to our sample and tests of the resulting correlations.  A discussion of the results is given in Section 6 and our conclusions are presented in Section 7.           

\section{ACCRETION THEORY PREDICTIONS}

\citet{JK99} examined the magnetspheric accretion theories of \citet{kon91}, \citet{camcamp93}, and \citet{shu94} deriving equations for the stellar magnetic field ($B_*$) in terms of $R_*$, $P_{rot}$, $\dot{M}$, and $M_*$ where $R_*$ is the stellar radius, $M_*$ is the stellar mass, $P_{rot}$ is the rotation period of the star, and $\dot{M}$ is the mass accretion rate onto the star.  One of the main assumptions underlying these accretion theories, and, as a result, the derived relationships, is that the central star is magnetically locked to its disk so that the star rotates at the Keplerian velocity of the disk truncation radius \citep{kon91,camcamp93,shu94,OS95}.  That is, the star and the inner truncation point in the disk are co--rotating.  As a result of torques acting through the magnetic field, the central star is spun down and is not able to rotate at break--up velocities, whereas high rotation rates would normally be expected if the star is allowed to contract and accrete without angular momentum being removed from the system.  By looking at the correlations predicted between stellar and accretion parameters by these magnetospheric accretion theories we can test the validity of these models as applied to PMS stars.  

Of the models examined, the \citet{kon91} and \citet{camcamp93} theories are purely mass accretion models and do not consider mass loss via a disk or stellar wind.  The \citet{shu94} and \citetalias{OS95}\footnote{OS95 is actually a detailing of the accretion flow and the role of flux pinching at the disk truncation point predicted by the \citet{shu94} model and not a new theory unto itself.  Thus the \citet{shu94} and \citetalias{OS95} references can be assumed to refer to the same model.} theories are different in that they combine accretion of mass onto the central star with the launching of a magnetocentrifugal disk wind from the disk truncation point.  The inclusion of wind--launching has important implications for the amount of matter that is ultimately accreted onto the star and also for the amount of magnetic flux that participates in the accretion flow.  The relationships predicted by \citet{kon91} and \citet{shu94} depend in the same way on the stellar and accretion parameters:

\begin{equation}
R_*^3 B_{dip} \propto
M_*^{5/6} \dot{M}^{1/2} P_{rot}^{7/6}.
\end{equation}    

\noindent The \citet{camcamp93} prediction deviates only slightly from eq. (1):

\begin{equation}
R_*^3 B_{dip} \propto
M_*^{2/3} \dot{M}^{23/40} P_{rot}^{29/24}. 
\end{equation}

\noindent In eqs. (1) and (2), $B_{dip}$ is the equatorial dipolar field strength at the stellar surface.  Plots created with eqs. (1) and (2) produce nearly identical results \citep{JK02}.  For this reason we only display plots generated using eq. (1) when considering the \citet{kon91}, \citet{camcamp93}, and \citet{shu94} models.

Standard magnetospheric accretion models all assume that the stellar magnetic field is purely dipolar.  Observations of magnetically sensitive absorption lines in TTSs, however, have shown that most TTSs have complex surface field geometries \citep{JK99,valenti04,daou06,donati08,donati11a,donati11b,donati11c}.  \citetalias{JK02} show that if the dipolar field requirement is relaxed, but that the amount of magnetic flux participating in the accretion flow is conserved and mapped back to the stellar surface from the truncation point (this is equal to one--third of the total magnetic flux trapped at the truncation radius; see \citetalias{OS95} for details), then the \citetalias{OS95} model can be modified to predict the following relationship:

\begin{equation}
R_*^2 f_{acc} B_{acc} \propto
M_*^{1/2} \dot{M}^{1/2} P_{rot}^{1/2} 
\end{equation} 

\noindent where \textit{f$_{acc}$} is the filling factor of the accretion columns on the stellar surface, i.e. the percentage of the surface covered by accretion flows.  In eq. (3) $B_{acc}$ is no longer the dipolar field strength but rather the strength of the field participating in the accretion flow at the surface of the star.  The magnetic flux participating in the accretion flow is equal to one--third of the total magnetic flux trapped at the truncation point \citep{OS95}.  Equation (3) (eq. 7 from \citetalias{JK02}) is a specific prediction of \citetalias{JK02}'s modified analysis of the \citetalias{OS95} theory of magnetospheric accretion.  More recently, investigations into the consequences of multipole magnetic field configurations on the observational properties of accretions flows have been undertaken using numerical models \citep[e.g.][]{mohanty08,long08,adams12}, providing further evidence that surface magnetic fields on TTSs are probably not purely dipolar.  However, \citet{adams12} find that contributions from higher order magnetic field components are negligible at typical disk truncation radii in TTS systems.  Thus the assumption that the dipole component of the magnetic field dominates when determining the truncation radius remains valid.

Measurements of mean surface magnetic fields on CTTSs have shown the field strengths to be relatively constant from star to star \citep{JK99,guenther99,valenti04,JK07,yangJK11}.  \citet{donati11b} have suggested, however, that the strength of the dipole and octopole components of the magnetic field on TTSs can vary strongly from star to star depending upon the existence and extent of a star's radiative core: fully convective stars host strong, stable dipolar fields \citep[e.g.][]{donati08,donati10} while stars with small radiative cores (0.0 $<$$M_{core}$/$M_*$$\lesssim$ 0.4) tend to have weaker dipolar fields and more dominant octopolar field components \citep{donati11a,donati11b}.  For example, \citet{donati08,donati10} measure 1.2 and 2--3 kG dipolar magnetic field components on the fully convective CTTSs BP Tau and AA Tau, respectively; for the partially convective CTTSs V2129 Oph and V4046 Sgr, however, \citet{donati11a} measure 900 and 50--100 G dipolar magnetic field components, respectively, for each star.  Although the number of stars for which individual components of the magnetic field have been measured is relatively small, the Donati et al. results suggest that these field components vary from star to star.  As mentioned above, the strength of the dipolar component has important implications for the location of the disk truncation radius.  If the dipolar field is weak enough, the accretion disk may pinch the field closer to the star allowing disk material to accrete along more complex regions of the star's magnetosphere \citep{romanova11}.  Such a scenario would most likely not be described by the equlibrium relationships of eqs. (1) and (3), due in large part to the fact that disk--locking can no longer be assumed if the truncation radius is much smaller than the co--rotation radius.


Equations (1) and (3) are directly testable if estimates for the necessary parameters can be obtained.  We note that $f_{acc}$ is not the same as the filling factor of the magnetic field on the surface of the star, which is taken to be 1.0 in the models being tested.  Estimates for $f_{acc}$, which are significantly less than one, can be obtained by modeling the accretion flow as a hot slab of emitting hydrogen \citep{valenti93,phart91}.  This procedure will be discussed in \S 4.  Due to the lack of magnetic field measurements for stars in NGC 2264, we are not able to substitute measured values of the surface magnetic field into eqs. (1) and (3).  We instead proceed by first making the simple assumption that the dipolar component of the magnetic field (eq. 1) and the magnetic field participating in the accretion flow (eq. 3) are constant from star to star.  Because we are concerned only with how well the two sides of eqs. (1) and (3) are correlated, assuming a constant field strength allows us to ignore $B_{dip}$ and $B_{acc}$ in the analysis.  While this assumption is justified based on measurements of mean surface magnetic fields on TTSs (see above), recent work by Donati et al. showing variations in the dipolar field component from one TTS to another potentially renders the constant field assumption inadequate, at least for stars which have developed radiative cores.  In an attempt to roughly mimic this behavior, we use the internal structure calculations from the \citet{siess00} PMS evolutionary models to determine the extent of the radiative cores in our sample.  We then assign an equatorial dipolar magnetic field strength based on the (limited) literature results in order to test the full forms of eqs. (1) and (3).  This procedure and the results of both methods (assuming a constant field vs. assigning individual field values) are presented in \S 5.        

\section{OBSERVATIONS AND DATA REDUCTION}

Our sample consists of 36 pre--main sequence stars of spectral type K and M in the open cluster NGC 2264 ($\alpha$=06:41, $\delta$=+09 35), and 14 main sequence dwarfs ranging from spectral type K1 to M4 which serve as spectral templates.  The sample is a subset of objects from the rotation study of \citetalias{lamm04}.  The stellar identification numbers are those used by \citetalias{lamm04}.  In the end, only eight of the observed templates are actually used in our models.  The NGC 2264 targets were chosen based on the availability of a determined spectral type and rotation period, and their confirmed PMS evolutionary nature based on approximate H$\alpha$ equivalent widths.  All of the stars in our sample have an H$\alpha$--index (a measure of the strength of H$\alpha$ emission) greater than .1, which indicates the stars are likely CTTSs \citepalias{lamm05}.  We briefly discuss the \citetalias{lamm04} classification of our sample in \S 3.3.  The rotation periods were determined based on photometric variability studied by L04; the spectral types were determined to within 1 subtype using optical and near infrared spectroscopy by \citet{rebull02}.  All of the stars in our sample are brighter than I$<$16.0.

\subsection{Observations}

A summary of the observing runs is given in Table 1.  We obtained low resolution (R$\sim$600) optical (3000 \AA -- 6000 \AA) spectrophotometry of each of the stars listed in Tables 2 and 3.  The wavelength resolution of the McDonald observations is $\sim$7.2 \AA; for the Kitt Peak observations $\Delta$$\lambda$$\sim$3.2 \AA.  Spectrophotometric flux standards taken from \citet{hamuy92} and \citet{oke90} were observed each night at various airmasses in order to generate atmospheric extinction curves and flux conversion factors.  Typical exposure times for the McDonald sample were $\sim$1000 seconds with a signal to noise ratio of $\sim$90--100 at 4400 \AA, approximately the center of the modeled wavelength region and free of any significant emission lines.  Exposure times for the Kitt Peak sample were typically $\sim$1200--1800 seconds with a signal to noise ratio of $\sim$40--50 at 4400 \AA.  Signal--to--noise ratios for fainter Kitt Peak stars (I$>$15) are closer to $\sim$20.  Typically the S/N ratio of the data decreases towards bluer wavelengths as the detector becomes less sensitive to these photons.        

On January 29, 2004 light cirrus clouds were present during the observations of HD10476, L5575, 5967, 6039, and 5394.  Exposures of L5967 and 6039 were retaken on Jan. 30 in order to obtain more accurate flux measurements.  L5355, 1316, 3636, and 6024 were also observed through light cirrus on March 1, 2005.  Calculated flux values for HD10476, L5575, 5355, 1316, 3636, and 6024 should therefore be taken as lower boundaries to the actual stellar flux.  This uncertainty is propagated through our model calculations and the final derived parameters for these stars are more uncertain than those derived for the stars with observations taken on photometric nights.  These values should be viewed with caution.   

\subsection{Data Reduction}

All of the data were reduced using custom IDL routines.  Bias and flat lamp images were taken at the beginning of each night.  For the McDonald sample, argon comparison lamp spectra were taken at regular periods during the night in order to account for wavelength shifts along the detector; for the Kitt Peak sample, iron-argon lamp spectra were used.  Cubic and quadratic dispersion solutions were calculated for each lamp spectrum using $\sim$15 lines from 3300\AA--5500\AA.  The lamp spectrum taken closest in time to the stellar exposure was used as the wavelength solution for that particular star.  Median bias images were constructed and subtracted from each science image.  In order to remove slit illumination effects by the flat lamp from the McDonald sample, polynomial fits were calculated perpendicular to the dispersion direction and then divided through the specified row in the flat image.  These normalized flats were then combined to a create a master median flat for each night.  The normalized flat was divided into each science image in order to remove pixel--to--pixel variations on the CCD.  Bad pixels and cosmic rays were removed manually by averaging over the adjacent two pixels in the dispersion direction.  Each science image was then corrected for atmospheric extinction and multiplied by the flux conversion factor calculated for that night.  The CCD read noise and statistical noise were included in the uncertainty calculations of the flux for each star.

\subsection{Grouping the Data}

Typical spectra and model fits are shown in Figure 1 (see Appendix A for the full sample), where the flux is given in units of $10^{-16}$ erg cm$^{-2}$ s$^{-1}$ \AA$^{-1}$.  The observed spectrum is shown by the solid black line; the fit to the data is overplotted with a red line.  The excess emission from the slab (dashed blue line) and the underlying scaled template (solid green line) are also plotted.  Our sample is separated into three categories: strongly accreting stars, weak accretors, and negligible accretors.  The negligible accretors (NA) were identified due to the lack of any excess emission required to produce good fits to the data, i.e. only a scaled and reddened template star was needed to match the observed spectrum.  The inclusion of a slab in these cases yielded accretion rates of $<$ $10^{-10}$ M$_\odot$ yr$^{-1}$, almost two orders of magnitude lower than the average rate for the strongly accreting (SA) stars, and changes in the $\chi^2$ goodness--of--fit measure of $<$10\%.  Model fits of the NAs are not shown in Figure 1.  The weak accretors (WA) display negligible Balmer jumps but required a slab in order to reproduce the observed flux blueward of the jump.  Chi--sqaured values for the WAs changed by $>$10\% when a slab was included in the model, indicating that low levels of mass accretion are occurring and producing a small amount  of excess emission.  Figure 1 clearly shows that the slab emission is a larger percentage of the total emission for the SA stars than for the WA stars.  Most SA stars show clear Balmer jumps and exhibit strong Balmer emission lines, as well as strong Ca {\sc ii} lines and some Fe {\sc ii} emission.  Fits to the strongly accreting sample changed significantly ($\Delta$$\chi^2$$>$50\%) when a slab was included in the model.  The average accretion rate for the SA stars is 10 times larger than the average rate for the WA stars.

The distinction between strongly accreting stars and weak accretors is defined here quantitatively by the ratio of the mass accretion rate to the stellar mass: if $\dot{M}$/$M_*$ $\geq$ 1 (in units of 10$^{-8}$ yr$^{-1}$) the star is categorized as a strongly accreting star; if $\dot{M}$/$M_*$ $<$ 1, the star is grouped as a weak accretor.  While this is an arbitrary distinction, the groups are useful when examining the relationships predicted by eqs. (1) and (3).  In addition, this criterion appears to separate the sample nicely into rough evolutionary regimes, with the majority of the SA stars having ages $<$ 5 Myr and the majority of weakly accreting stars having ages $>$ 5 Myr.  Age determinations for the sample are discussed in \S 4.4.  The negligible accretors display a range of ages and these stars are most likely WTTSs whose disks have largely dissipated.  By determining which of the stars belong to the NA category we are able to eliminate objects that might yield unrealistic slab and accretion parameters due to their very low or non--existent accretion rates.  The disk--locking theories being tested here most likely no longer apply to these stars. 

The sample was chosen based on the \citetalias{lamm05} classification of almost our entire sample as CTTSs according to a definition based on \textit{R}--band and H$\alpha$ magnitudes.  This definition is similar to the original classification of the stars based on H$\alpha$ equivalent widths, with CTTSs having W(H$\alpha$) $\geq$ 5 \AA$\;$\citep{bert89}.  However, it is obvious from the spectra of the negligibly accreting stars (Figure 1, panel c) that these stars, and a small number of the weakly accreting stars, do not show typical characteristics of CTTSs, e.g. strong Balmer emission lines and large veilings at blue wavelengths.  \citetalias{lamm05} state that this classification definition selects stars that ``most likely show the properties of accreting PMS stars'' but that it does not provide strong proof of their inclusion in one group or another, due to the potentially large equivalent widths of H$\alpha$ from chromospheric emission.  Thus, we find that our SA and WA stars are probably CTTSs, while our NA stars are most likely WTTSs.

\section{MODELING THE EXCESS}

The excess emission from CTTSs is believed to result from an accretion shock at the stellar photosphere: matter accreting along magnetic field lines slams into the stellar surface at near free--fall velocities, is strongly shocked and heated, and emits in the X--ray, UV, and optical \citep[e.g.][]{kenyon94,kastner02}.  This excess emission is believed to be produced over a range of temperatures and densities along the shock column.  The picture becomes more complicated if one considers the heating of the underlying photosphere, spot effects, and the interplay between the pre--shock and post--shock region.  While simple 1--D shock codes have been utilized as a more realistic method to produce the excess spectrum \citep[e.g.][]{calvet98,muz03}, uncertainties in the temperature structure of the accretion flows and how they are heated remain unresolved issues for more advanced shock models.  For our purposes, however, all that is needed is a reliable estimate of the excess accretion emission, distinct from the stellar photospheric emission.  

Good estimates of the excess emission can be obtained by treating the accretion flow as a single--temperature accretion column, or slab, in LTE originating at the photosphere of the star \citep{valenti93,gullbring98}.  Estimates derived from slab models show good agreement with those calculated from shock models \citep{calvet98} and slab models continue to be used to measure accretion rates on young objects \citep[e.g.][]{herczeg08,herczeg09}.  Although this is almost certainly not the physical scenario producing the excess emission, order of magnitude estimates of the emission are good enough to the test relationships predicted by the accretion theories being investigated here.  In any case, systematic uncertainties in the mass accretion rate may dominate compared to differences produced by differences in the model treatment of the accretion flow itself, at least as far as slab and 1--D shock models are concerned.  We discuss some of these systematic uncertainties in \S 4.3.   

\subsection{Slab Models}

We model our sample of PMS stars by combining a scaled main sequence template star spectrum with the scaled slab emission which represents the emission from the accretion column.  The parameters governing the slab emission are number density, \textit{n}, temperature, \textit{T}, the length, \textit{l} through the slab, and the filling factor, $f_{acc}$, of the total stellar surface area covered by the accretion columns.  The emitting slab in the modeled region (3400 \AA -- 5500 \AA) is taken to be composed of pure hydrogen.  \citet{phart91} found that metals contribute $<$5\% of the electrons in the slab for typical temperatures and densities in our models.  In addition, \citet{phart89} show that emission line flux from all elements constitutes $<$5\% of the total excess flux.  For our purposes here we are primarily interested in estimating the accretion luminosity and resulting accretion rate, and uncertainties associated with the exclusion of elemental emission lines are acceptable since the uncertainties in the accretion rate are likely dominated by larger uncertainties, such as degeneracies in the fits themselves, geometric factors, and uncertainties in the distance and stellar radii.  Thus, emission lines from other elements are not included in the model calculations.  We also fix the turbulent velocity of the slab gas at 150 km/s in order to roughly match the line widths seen in our sample.  The turbulent velocity actually serves as a substitute for the wavelength resolution of the instrument and does not represent a physical property of the slab.  Letting the turbulent velocity vary in our fits does not result in significant improvements to the fit.  If the line width is not well matched by $V_T$=150 km/s then $V_T$ is manually adjusted until a good fit to the line width is achieved.  

The slab emission is calculated by solving the Boltzmann and Saha equations for the given temperature and density and then evaluating the optical depth and source function, in this case the Planck function, at each wavelength.  The maximum number of levels included in the hydrogen atom (typically $\sim$80) is determined by the temperature and density of the slab.  The total opacity of the slab is calculated using the sum of H{\sc i}, H{\sc ii}, H$^{-}$, and electron scattering opacities.  The most important feature of the slab is the Balmer jump, which occurs at $\sim$3650 \AA $\;$and we define here to be ($f_B$-$f_R$)/$f_R$, the flux just blueward of the jump minus the flux just redward of the jump normalized to the red continuum \citep{valenti93}.  Our low resolution spectra do not resolve the tightly packed higher Balmer lines, which merge into a pseudo--continuum around 3700 \AA.  Thus, the Balmer lines blueward of H$\delta$ are ignored in our fitting procedure.  H$\gamma$ and H$\beta$ are also ignored when fitting the data due to the possibility of winds significantly contributing to the line strengths \citep{alencar00}.  For this reason we have chosen to use only H$\delta$ as a constraint on the Balmer line strengths.  This generally produces good fits to the higher Balmer lines but poorer fits to H$\gamma$ and H$\beta$.  However, we are mainly concerned with fitting the excess continuum and thus are not overly concerned with emission models that produce the best fits to the lines.  To calculate the emergent spectrum, we solve the radiative transfer equation through the slab.                      

Our model procedure attempts to match the observed spectrum, particularly the Balmer jump.  It is valuable to discuss how variations in the slab parameters affect the shape of the slab emission.  The size of the Balmer jump is governed by the temperature of the gas and the optical depth through the slab.  Increasing the temperature for a fixed density results in a weaker jump due to a faster increase in the populations of levels n$>$2 compared to increases in the n=2 population.  In other words, the optical depth blueward of the jump decreases relative to the optical depth redward of the jump and the jump is diminished.  The optical depth through the slab is controlled by \textit{n}, \textit{l}, and, to a lesser extent, \textit{T}.  Typical values for \textit{n} and \textit{l} that we find in our models are $10^{14}$ cm$^{-3}$ and $10^7$ cm, respectively.  For these values and a slab temperature of 9000 K, the slab continuum is optically thin ($\tau$ $<$ 1) and the Balmer lines are very strong, i.e. $\tau$$>$$>$1 in the lines.  As the density is increased, the slab becomes optically thick and the line emission is weakened relative to the increasing continuum emission.  For densities above $10^{15}$ cm$^{-3}$, the slab becomes optically thick and is dominated by H$^-$ continuum emission.  Increasing \textit{l} will also increase $\tau$.  Because our model varies \textit{n}, \textit{l}, and \textit{T} simultaneously, similar optical depths through the slab can be obtained for varying combinations of the parameters.  However, we are more concerned with their product, $\tau$, than with the specific values of the parameters themselves.  This interrelation between model parameters is discussed more fully in Appendix B (see Tables B1 and B2 and Figures B1 and B2).  A detailed discussion of our fitting procedure and the assumptions included in our model are also given in the appendix.     

\subsection{Main Sequence vs. Pre--main Sequence Templates}

It is well known that WTTS, though not accreting, still have active chromospheres capable of producing small amounts of excess emission at UV wavelengths \citep[e.g.][]{houdebine}.  \citet{ingleby} recently pointed out the advantage of using WTTS templates in fits to CTTS spectra in order to determine more accurate accretion excesses; use of MS templates can result in an over--estimate of the accretion luminosity, $L_{acc}$.  In order to test the impact of using MS templates for our sample, we have modeled 8 stars using the WTTS template V830 Tau.  V830 Tau is a K7 star so all of the modeled stars in our sample are those that required a K7 MS template in the original fits.  Before using V830 Tau as a template it was de--reddened and convolved with the instrumental resolution of the McDonald and Kitt Peak intsruments to match the spectral resolution of our observations.  A comparison of the results is shown in Table 4.

It is obvious that the parameters generated by the two separate fits are not exactly the same.  Overall, however, there is general agreement.  The most important result pertains to the accretion rate estimates, for which both template stars produce similar values.  Higher $\dot{M}$ values from MS templates are expected based on the lack of any excess emission from active chromospheres \citep{ingleby}.  This behavior is weakly reflected in our comparison where the MS fits produce $\dot{M}$'s that are larger on average by a factor of 1.13.  In any case, the small differences found here do not affect the comparisons discussed in \S 5.  We conclude that, for the purposes of this study, fitting with MS templates as opposed to PMS templates results in negligible changes to the relevant parameters.   

\subsection{Determining the Stellar and Accretion Parameters}


Once $R_*$ is determined from the model, the luminosity of the star can be determined by adopting the effective temperature corresponding to the spectral type of the best--fit template star using the spectral type--$T_{eff}$ calibrations discussed in Appendix A.  The stellar luminosity and effective temperature can then be used to estimate the stellar mass and age by placing the star on an HR diagram and comparing its position with pre--main sequence evolutionary tracks \citep[e.g.][]{siess00,palla99,baraffe98}.  We find good agreement between the masses determined using the \citet{siess00} tracks and those of \citet{palla99}.  Masses determined from the \citet{baraffe98} tracks typically differ from the other two by 50\%.  This is not unexpected due to differences in the way each set of models treats the stellar interior equation of state and convective energy transport \citep{siess00,hillenbrand04}.  The ages, however, determined using \citet{baraffe98} tracks are systematically higher by a factor of $\sim$2-10 than those found using the \citet{palla99} and \citet{siess00} models.  The large age difference produced by using different sets of pre--main sequence eveolutionary models is important to keep in mind when comparing cluster ages determined by different studies.

\citet{hillenbrand04} find that PMS models do not agree well with dynamical mass estimates for M$_*$$<$0.5 M$_\odot$ $\;$and that the models underpredict actual masses by 10\%-30\% for 0.3 M$_\odot$$<$M$_*$$<$1.2 M$_\odot$.  Again, the uncertainty is mainly due to our relatively poor knowledge of convective energy transport and the necessary opacities at low effective temperatures.  A large portion of our sample (80\%) have estimated masses $<$ 1.2 M$_\odot$, while $\sim$33\% have masses $<$ 0.5 M$_\odot$.  At these low masses the models become even more uncertain, due in part to the uncertain location of the stellar birthline \citep{baraffe02}.  For the remainder of this paper, we choose to use the \citet{siess00} models in order to determine mass and age estimates for our sample, though use of the quantities determined from the \citet{palla99} models produces only small differences ($<$10\%) in the final stellar and accretion parameter values (e.g. $R_*$, $\dot{M}$, and $f_{acc}$).  More specifically, we use the Z=0.01 (convective overshooting not included) tracks for our mass and age estimates.  \citet{james06} found some evidence that young clusters tend to be metal--poor, although the number of clusters examined in their study (3) is small.  Thus, we adopt the slightly less--than solar metallicity tracks for our analysis.  In any case, \citet{mayne07} have shown that age estimates for young clusters do not differ significantly based on the adopted metallicity in the evolutionary model and a comparison of the \citet{siess00} Z=0.01 and Z=0.02 (approximately solar metallicity) tracks shows that differences in mass estimates are negligible at comparable values of $L_*$ and $T_{eff}$.

Figure 2 shows our sample plotted on the HR diagram.  Tables 5, 6, and 7 list $M_*$, the age, $R_*$, $T_{eff}$, $L_*$, the rotation period $P_{rot}$, \textit{f$_{acc}$}, $A_V$, and the derived accretion rate, $\dot{M}$, for each star.  Accretion rates and filling factors are not calculated for the NA stars and so are not included in Table 7.  Uncertainties in the mass and age estimates, apart from the inherent uncertainties in PMS evolutionary models, are dominated by uncertainties in the underlying effective temperature of the photosphere and also by uncertainties in the stellar radius, which is determined from the model.  Typical error bars for $L_*$ and $T_{eff}$ are shown on Fig. 2.  The uncertainties in $T_{eff}$ and $L_*$ translate into 30-50\% uncertainties in the stellar ages and 20--25\% uncertainties in $M_*$.  We also note that the radiative core masses used in Sections 5 and 6 are subject to similar, if not larger, uncertainties as $M_*$. 

Good fits to the stellar continuum should provide reliable estimates of the underlying contribution of the photosphere.  If the underlying stellar contribution is known, and $T_{eff}$ is estimated with some confidence, the radius can be determined with typical uncertainties of $\sigma_R$=$\pm$15\%.  Once $R_*$ and $M_*$ have been determined, $\dot{M}$ can be directly calculated using the following relationship \citep{gullbring98}:

\begin{equation}
\dot{M}=\frac{L_{acc}R_*}{GM_*}\left(1-\frac{R_*}{R_{Tr}}\right)^{-1}.
\end{equation}

\noindent To determine $\dot{M}$ we take the disk truncation radius, $R_{Tr}$, to be 5$R_*$.  The accretion luminosity, $L_{acc}$, is determined by adding up the slab flux $F_{acc}$ across the wavelength range 100 $\leq$ $\lambda$ $\leq$ 40000 \AA$\:$and applying the relationship

\begin{equation}
L_{acc}=4\pi R_*^2f_{acc}F_{acc}
\end{equation}

\noindent which only considers the luminosity of the accretion columns onto the stellar surface.  The flux, $F_{acc}$, is primarily determined by the observations, i.e. from the strength of the Balmer jump and the shape of the Paschen and Balmer continuums.

It can be seen from Figure 2 that the SA stars, which have an average age of $\sim$2.0 Myrs, appear generally younger than the WA and NA stars.  The WA stars ($\sim$10.6 Myrs) are on--average similar in age to the NA stars ($\sim$9.2 Myrs).  The NAs show a large spread in age and rotational period.  Large age spreads in star forming regions are common \citep[e.g.][]{dahm05,hillenbrand97,kenyon95} and the approximate age of the cluster is usually taken to be the most common value in the distribution.  Due to the small number of stars in the NA group, we cannot make any definitive statements about their evolutionary status.  We can, however, generalize and say that the older, shorter period NA stars have probably decoupled from their disks and begun to spin up as they contract.  L4511 ($\sim$7 Mrys) and L6024 ($\sim$1 Myr), both longer period stars, have most likely decoupled from their disks only recently and have not had time to spin--up appreciably.  L4098, which is both young and fast--rotating, seems to have not had time to establish disk--locking and thus is still contracting without being regulated by its disk.

The clear separation in age of the SA and WA/NA stars hints at a real age difference between the groups.  This same behavior is seen in the Taurus--Auriga star forming region: \citet{bert07}, using revised distance estimates for 36 WTTSs and 30 CTTSs, show that the CTTSs are systematically younger as a group than the WTTS population and thus the former are most likely the predecessors of the latter.  The result from \citet{bert07} is the first convincing observational evidence that CTTS and WTTS are not coevolutionary but rather WTTS are the evolved products of the CTTS phase.  Our HR diagram of NGC 2264 (Fig. 2) seems to support this scenario, although our sample size is approximately half that of the \citet{bert07} study.

Our sample displays a large spread in age, with some stars seemingly as old as 50 Myrs.  The average age is $\sim$6.4 Myrs which does not agree very well with previous, more statistically significant age determinations of the cluster (Mayne et al. 2007--3 Myrs; Dahm \& Simon 2005--1.1 Myrs for the TTS population; Rebull et al. 2002--3 Myrs; Park et al. 2000--3.2 Myrs; Sung et al. 1997--3 Myrs; Walker 1977--3 Myrs).  To our knowledge, our study and that of \citet{mayne07} are the only ones that employ the \citet{siess00} models to derive age estimates for targets in NGC 2264\footnote{\citet{rebull02} compare the ages obtained for NGC 2264 using the \citet{siess00} models with those of \citet{dantona94} but they choose to adopt values from the latter models.} so it is important to reiterate that different PMS evolutionary models can produce very different age estimates for the same object.  As we noted above, comparisons between ages determined from the \citet{siess00} models and the \citet{baraffe98} PMS tracks yield values that differ by up to an order of magnitude; individual ages should be examined with caution.

\subsection{$L_{H\beta}$ vs. $L_{acc}$}

We have measured the H$\beta$ line luminosities for our sample.  The line flux was calculated by subtracting the continuum from 4820--4840 \AA$\;$averaged with the continuum from 4880--4900 \AA$\;$and summing up the residuals from 4840--4880 \AA.  A distance of 760 pc is assumed when computing the line luminosities (see Appendix B for a discussion of distance estimates).  The $A_V$ values from Tables 5--7 were used to correct each star for extinction.  We used our SA and WA stars to determined the relationship between the H$\beta$ line luminosity and the accretion luminosity.  Figure 3 shows a strong relationship between the accretion luminosity, $L_{acc}$, and the H$\beta$ line luminosity, $L_{H\beta}$.  The best--fit relationship to the data is

\begin{equation}
\log\left(\frac{L_{acc}}{\Lsun}\right) = (2.80\pm.21)+(1.12\pm.06)\log\left(\frac{L_{H\beta}}{\Lsun}\right).
\end{equation}

Equation (6) is almost identical to the relationship calculated by \citet{fang09} for TTSs in the Taurus--Auriga star forming region.  Eq. (6), however, was calculated by excluding any stars with log$_{10}$($L_{H\beta}$/$L_\odot$) $<$ -4.50 in order to account for chromospheric line emission not produced in the accretion flow.  This number was chosen by averaging the H$\beta$ line luminosities of 5 WTTSs in Taurus (Table 8), each of which show no evidence of a circumstellar disk at near--IR or millimeter wavelengths and therefore are most likely experiencing negligible accretion.  Below this threshold the H$\beta$ luminosity is likely dominated by chromospheric emission.  Therefore, it is not possible to measure the H$\beta$ luminosity produced by accretion for TTSs that lie below this value and are within the mass and radius range studied here, i.e. H$\beta$ cannot be used as an accretion proxy for these stars.  The NGC 2264 objects with line luminosities below this value are not included in the determination of eq. (6).  Also plotted in Fig. 3 are the NA stars (red diamonds) with measureable H$\beta$ emission.  They are plotted using the accretion luminosities that result from fitting the spectrum with a template and a slab.  All of these objects are clustered near the threshold and lie below the line of best fit (eq. 6), thus indicating that H$\beta$ line strength should not be used as the sole proxy for the accretion luminosity.  The NA stars are not included when determining eq. (6).     

The \citet{fang09} relationship includes $\sim$7 objects below our estimated threshold.  We do not know the identities of these objects\footnote{\citet{fang09} take their H$\beta$ luminosities from the studies of \citet{herczeg08} and \citet{gullbring98}.} but they are most likely late M--type stars and brown dwarfs from \citet{herczeg08} in which case our threshold estimate may not apply if the chromospheric contribution to H$\beta$ decreases at very low masses.  The application of eq. (6), however, to a large sample of TTSs \textit{in the same mass range as our sample and with $L_{H\beta}$ less than the threshold value} is cautioned against due to the intrinsic H$\beta$ luminosity produced by active chromospheres.  Doing so can lead to unrealistic accretion luminosity estimates and in turn to incorrect mass accretion rates.

\subsection{Comparison with Rebull et al. (2002) Values}\defcitealias{rebull02}{R02}

\citet{rebull02} (hereafter R02) computed the age, $M_*$, $R_*$, $A_V$, and $\dot{M}$ for 16\footnote{Values of $A_V$ and $\dot{M}$ are not calculated by R02 for a few individual stars in this subsample.} of the 36 stars in our sample.  A comparison of the values is given in Table 9.  Visual comparisons of $A_V$ and $\dot{M}$ are plotted in Figures 4 and 5, respectively.  Differences in the derived masses range up to 75\% but, on average, they agree to within 30\%.  The most striking difference between the parameters is the age determinations.  The large disparity can be primarily attributed to the use of different PMS evolotionary models.  \citet{rebull02} use the \citet{dantona94} tracks while we use the \citet{siess00} tracks.  As \citetalias{rebull02} point out (\citetalias{rebull02}, Fig. 13), the \citet{dantona94} models consistently produce ages younger than the \citet{siess00} models by .5--dex.  

The $A_V$ values derived by \citetalias{rebull02} for several stars using ($R$--$I$) color excesses are significantly different than our $A_V$ values, which are the best--fit values from our models (Figure 4).  We have attempted to model our stars using the $A_V$ values from \citetalias{rebull02}.  If $A_V$ is fixed at the \citetalias{rebull02} value, however, the fit is either not of comparable quality to the fit determined by letting $A_V$ vary or our routine fails to find an acceptable match to the data.  For a few of the stars, the difference in $A_V$ can be attributed to differences in the spectral type determined by \citetalias{rebull02} and the spectral type of the template star used to model the spectrum.  In order to derive accretion rate estimates, \citetalias{rebull02} use \textit{U}--band magnitudes to estimate the accretion luminosity based on the relationship from \citet{gullbring98}.  They are well correlated (Figure 5), with \citetalias{rebull02} producing accretion rates higher on average by a factor of $\sim$2.5, though the factor is larger for stars with smaller accretion rates and smaller for stars with higher $\dot{M}$.

There is one discrepency, however, that deserves closer attention.  \citet{rebull02} calculate a large accretion rate ($\dot{M}$=1.3 x 10$^{-7}$ M$_\odot$ yr$^{-1}$) for the star L6172, while we determine, based on L6172's optical spectrum, that it is not significantly accreting.  Using only a scaled and reddened template star, we find an excellent fit to the L6172 data.  If we allow some accretion luminosity to contribute to the spectrum, fits of comparable quality can be obtained which produce very low accretion rates of $<$ 8 x 10$^{-10}$ M$_\odot$ yr$^{-1}$.  A glance at Fig. 1c confirms the lack of any obvious excess emission redward of 3400 \AA.  Fig. 1c also shows that L6172 has reasonably strong H$\beta$ line emission, but this could be due almost entirely to chromospheric activity.  Using eq. (6) and the values of $M_*$, $R_*$, and $\dot{M}$ from \citetalias{rebull02}, we have plotted L6172 in Fig. 3 (open blue circle).  The H$\beta$ luminosity calculated using eq. (6) is log$_{10}$($L_{H\beta}$/$L_\odot$)=-2.59, while we find log$_{10}$($L_{H\beta}$/$L_\odot$)=-3.76, more than an order of magnitude less than the value calculated using the \citetalias{rebull02} values.  Although TTSs are known to show significant variability on a variety of timescales \citep{bert89}, routine variability studies typically do not find variations in the accretion rate of 3 orders of magnitude.  For example, \citet{alencar02} find accretion rate variations for TW Hya of $\sim$1 order of magnitude over a year timescale.  In a UV variability study of BP Tau, \citet{ardila00} measure accretion rates that vary by a factor of 7 over $\sim$1 year timescales.  The FU Orionis phenomena can produce increases in the mass accretion rate by 4 orders of magnitude or more but the decay timescale is often longer than a decade \citep{herbig77,hartmann96}.  Our observations were taken $\sim$8 years after those of \citetalias{rebull02} so the possibility of L6172 having undergone an intense period of mass accretion during the latter set of observations cannot be entirely ruled out.  

Although it is not clear to us what the source of this discrepency is, based on the relative agreement of our other accretion rates with those of \citetalias{rebull02}, we have confidence that L6172 is not a heavy accretor in our data for 2 reasons: 1. It does not appear to have any excess emission in the optical, and 2. $P_{rot}$ for L6172 is 1.76 days which implies that it is most likely rotating freely and is not locked to its disk.  Although L6172 appears to be very young ($\tau_{L6172}$ $<$ 2 Myr in both studies), which makes it more unlikely that it has ceased interacting with its disk, it is certainly possible that it has yet to establish disk--locking, or is in the process of doing so.  This is consistent with the ``moderate'' angular momentum loss scenario used by \citetalias{lamm05} to explain the presence of young, fast--rotators with large IR excesses in Orion and NGC 2264.  We believe that L6172 falls into this category.  Aside from this individual case, we find good agreement within the (large) uncertainties between our parameters and those of \citetalias{rebull02}.

\subsection{A Note on Uncertainties in $\dot{M}$}

We have already discussed the uncertainties in our model fits (see Appendix B for further discussion).  Calculated values of $\dot{M}$ rely on the stellar and accretion parameters derived from the models.  Assumptions concerning the geometry of the system and the adopted value for $R_{Tr}$ introduce a factor of $\sim$2 uncertainty in individual accretion rates.  For example, we assume that the observer is looking straight ``down'' on the accretion column instead of modeling multiple accretion columns at different angles and projections onto the surface.  We also assume that half of the accretion luminosity is emitted into the solid angle subtended by the star, while half is emitted outward toward the observer.  These assumptions affect the final values of $L_{acc}$ and $f_{acc}$ which translate into uncertainties in $\dot{M}$.  A more obvious uncertainty arises in the choice of $R_{Tr}$ which can result in differences in $\dot{M}$ of a factor of $\sim$1.6 depending on the adopted value in the range 3--6 $R_*$.  In addition to these uncertainties, we can estimate uncertainties in $\dot{M}$, within the context of our assumptions concerning the accretion flow and system geometry, once we estimate uncertainties in the derived parameters.

Our $\chi^2$ fitting routine estimates uncertainties in the model parameters based on the quality of the fit.  We can easily translate these uncertainties into uncertainties for $R_*$ and $M_*$.  If we assume an uncertainty of 150 K in $T_{eff}$, uncertainties in $M_*$ are generally near 20--25\%.  Typical uncertainties for $R_*$ are 15\%.  For accreting stars, $L_{acc}$ is relatively well constrained by shape of the spectrum and its measured strength, although $R_*$ and $f_{acc}$ are used to calculate the final value.  To compute the uncertainty in $L_{acc}$, we assume that the uncertainty in $F_{acc}$ is equal to the errors in the observed spectrum.

It is important to highlight the role that $A_V$, the visual extinction, plays in determining the accretion rate.  Higher values of $A_V$ will cause the bluer portion of the spectrum to be enhanced relative to the red wavelengths thus essentially increasing the slab luminosity required to generate a good fit to the spectrum.  A higher accretion luminosity will result in a higher value of $\dot{M}$ assuming that $R_*$ and $M_*$ remain the same.  As we demonstrate in Appendix B, we are relatively confident in the best--fit parameters for each spectrum, i.e. the spread in the final values of each parameter is small when the input parameters are varied.  This is true for $A_V$, as can be seen in Table A2.  As mentioned in \S 4.5, however, our model $A_V$ values can differ significantly from those derived by \citetalias{rebull02} (Figure 4).  The disagreement between our model $A_V$ values and those determined directly from the \citetalias{rebull02} photometry are not surprising considering the different wavelength regions utilized in the calculations.  As we discussed in \S 4.5, fixing $A_V$ at the \citet{rebull02} values when fitting the spectra generally does not produce models of the quality obtained by letting $A_V$ vary.  We highlight these points simply to illustrate the caution necessary in considering any individual value as a precisely determined quantity.                

Combining all of the uncertainty estimates yields uncertainties in $\dot{M}$ of 35--50\%.  Thus, uncertainties in $\dot{M}$ produced by uncertainties in the stellar and accretion parameters are small relative to the uncertainties in the treatment of the accretion flow itself.  For this reason, individual values of $\dot{M}$ should be view with caution.  However, most of the systematic uncertainties, as mentioned previously, should not affect the final correlations presented in Section 5: that is, such uncertainties may shift the stars as a group, but it is unlikely that the errors that may be present will introduce a correlation when none is actually present.

\section{ANALYSIS AND RESULTS}   

In order to test the magnetospheric accretion models outlined in \S 2, we use the stellar and accretion parameters from Tables 5 and 6 to plot the left--hand side vs. the right--hand side of the relationships predicted by eqs. (1) and (3).  We then test whether or not the data are correlated.  The NA stars are not included in our analysis because they are not accreting measureable amounts of gas from their disks and thus are most likely no longer strongly coupled to their disks.  The same analysis is performed for both the assumption of a constant magnetic field from star to star and our assignment of magnetic field strengths to each individual star based on the size of the star's radiative core.

We have discussed the uncertainties in our derived parameters at some length.  Most of these uncertainties are systematic and dominate the observational uncertainties to a large degree.  For this reason, we expect the measured errors to produce scatter about the predicted relationships while the systematic uncertainties will tend to shift the relationships but not create correlations where none actually exist.  Scatter about the predicted relationship is probably also due to fluctuations in the equilibrium configuration of the CTTS system.  In reality, CTTS systems most likely experience variable accretion rates \citep[e.g.][]{ardila00,vanboekel} leading to variations in $\dot{M}$, $P_{rot}$, and $R_{Tr}$, the disk truncation radius.  Long--term variations in the strength of the dipolar component of the magnetic field \citep[e.g. V2129 Oph; see][]{donati07,donati11a} may also affect the equilibrium star--disk configuration \citep{donati11a,donati11b} which may in turn affect the correlation measurements.  Thus the correlations predicted by eqs. (1) and (3) are likely only valid for the average of these fluctuations around the equilibrium.  This effect can be seen in Fig. 8c where a strong correlation is present and also exhibits the expected scatter surrounding the line of best--fit.

In order to quantify the strength of the relationships given by eqs. (1) and (3) as applied to our sample, we compute the Spearman's rho rank correlation coefficient, \textit{r}, and the correlation significance $P$ \citep{press87} for three separate group of stars: the strong accretors, weak accretors, and the combination of strong and weak accretors.  These results are shown in Figures 6 (eq. 1) and 8 (eq. 3) for the case of a constant magnetic field for the sample.  We also tested eq. (1) using stars from \citetalias{rebull02} for which rotation periods are available from \citetalias{lamm04}.  This sample is shown in Fig. 7.  In order to prevent extreme values from overly influencing the correlations, we consider the logarithm of the quantities in the proportionalities.  The correlation significance, $P$, is a measure of how likely it is that the calculated correlation is not actually present in the data.  Thus, a lower value for $P$ implies a lower liklihood that the correlation is not real.  Values of $P$$<$.01 (99\% confidence level) are generally considered evidence of a real correlation.  Values of \textit{r} close to -1.0 and 1.0 imply strong correlations.  A summary of the correlations for each equation, their associated values of $P$, and the slope of the best--fit line are given in Table 10 for the case of a constant magnetic field strength from star to star.  Best--fit lines are not calculated for the poor correlations produced using eq. (1).

It is obvious from Table 10 and Figs. 6a, b, c, and 7 that only weak correlations, if any, exist for eq. (1), which is derived from the \citet{kon91} and \citet{shu94} magnetospheric accretion models by assuming a purely dipolar magnetic field and a stellar rotation rate equal to the Keplerian rate in the disk at $R_{Tr}$.  It is equally clear from Figs. 8a, b, and c that eq. (3), a relationship predicted by \citetalias{JK02}'s modified version of the \citetalias{OS95} theory to include a non--dipolar surface magnetic field, yields excellent correlations with low values of $P$ for all three samples.  Again, we have assumed a constant magnetic field for our sample when plotting the proportionalities in Figs. 6, 7, and 8.

We applied the same analysis to our sample after assigning magnetic field strengths to each star based on the extent of the star's radiative core, following the suggested evolution of the dipole by \citet{donati11a}.  The internal structure of each star was estimated using the \citet{siess00} models.  Using the current observational results of dipolar field strengths on TTSs as a guide \citep{donati11a,donati11b,donati10,donati08}, we chose values of 0.1, 0.5, and 1.5 kG to be assigned to stars with $M_{core}$/$M_*$$>$0.4, 0.0 $<$ $M_{core}$/$M_*$$\lesssim$ 0.4, and $M_{core}$/$M_*$=0.0, respectively, where a star with $M_{core}$/$M_*$=0.0 is fully convective.  In other words, instead of assuming a constant magnetic field from star--to--star we let it vary according to its internal structure to see the overall effect on the correlations.  We note that the values given above for the surface dipolar field are values at the magnetic pole, while eq. (1) uses the equatorial field strength.  For the dipolar field component, the equatorial value is one--half of the polar value.  Although eq. (3) involves the surface magnetic field value participating in the accretion flow, independent of its specific geometry, we have used the same field values given above as for the purely dipolar case of eq. (1).  After letting the magnetic field values vary, the correlations actually worsen for both equations: for eq. (1), we now have $r$=0.13 and $P$=0.52; for eq. (3), we now have $r$=0.75 and $P$=2x10$^{-5}$.  The variation of the magnetic field essentially spreads out the data along the abscissa and makes the best--fit slope shallower (slope=0.50) so it is not surprising that the correlations become weaker.  Thus, while the correlation strengths change when the dipole component of the magnetic field is allowed to vary from star--to--star, our results still show stronger support for the \citetalias{OS95} model than for the purely dipolar magnetospheric accretion models from eq. (1).  For the remainder of this paper we will refer to the constant magnetic field analysis for both eqs. (1) and (3) when discussing the implications of our results.    

The slope of the predicted relationship should be 1.0 using eqs. (1) and (3).  For the initial linear fit to the data, we assume that each point has an equal uncertainty.  The standard deviation of the residuals between the best--fit line and the data are then assumed to be the 1$\sigma$ uncertainties for each point.  These 1$\sigma$ values are then used to compute the uncertainty in the slope itself.  As can be seen in Figure 8c and Table 10, the slope (0.63$\pm$0.03) of the best--fit line to the data (solid line) does not match the predicted line (dashed) slope of 1, even to within 3$\sigma$.  \citetalias{JK02} also find a shallower slope (0.73$\pm$0.14) for their sample of stars in Taurus.

The cause of the slope differing from the predicted value in Taurus and NGC 2264 is currently unkown; however, one possibility may have to do with the evolution of the stellar magnetic field geometry.  As described in \S 2, \citet{donati11b}, building off the work of \citet{morin08}, suggest that very young, fully convective PMS stars have field geometries with relatively strong dipole components.  As they age and develop even a small radiative core, Donati et al. suggest the surface field becomes much more complex, with higher order terms becoming stronger than the dipole.  As a result, younger stars will have magnetic interactions with their disks at several stellar radii that are essentially the same as for a pure dipole as assumed by \citetalias{JK02}. However, for older stars, the higher order magnetic components may start to change the nature of the field interaction with the disk and thereby change the specific predictions of how stellar and accretion parameters should scale with one another.  This could result in a changing of the slope of the predicted relationships.  If true, the youngest clusters should show correlations with slopes close to 1.0 and the slope should evolve away from this value for older clusters.  This is essentially what is seen with \citetalias{JK02} finding a slope closer to 1.0 for their Taurus sample than we find for our slightly older NGC 2264 sample.

At first glance, our data appears to strongly support \citetalias{JK02}'s modified version of the \citetalias{OS95} theory.  Before we are able to interpret the strong correlations as conclusive support for the modified \citetalias{OS95} model, however, we must explore any interdependencies of the variables that may arise for other reasons.  In this way, we can ensure that the observed correlations are in fact evidence for the model predictions and are not driven by expected relationships between parameters based on how they are determined.

\subsection{Testing the Correlations}

For a uniform flow, the mass accretion rate is

\begin{equation}
\dot{M}=
4 \pi R_*^2f_{acc} \rho v
\end{equation}

\noindent where \textit{v} is the velocity of the flow and $\rho$ is the density, both measured at the point of impact on the stellar surface.  It is clear that $\dot{M}$ should be correlated with $R_*^2$$f_{acc}$.  Thus it is important to test whether or not $\dot{M}$ and $R_*^2$, $\dot{M}$ and \textit{f$_{acc}$}$R_*^2$, and $\dot{M}$ and $f_{acc}$, show any signs of being correlated in our sample.  Table 11 shows the results of these comparisons.  Clearly there is a significant correlation between $\dot{M}$ and $R_*^2$ (Figure 9) and a lack of correlation between $\dot{M}$ and $f_{acc}$.  In fact, one might worry that the correlation shown in Fig. 8c for the full sample of SA and WA stars plotted using the modified \citetalias{OS95} prediction is dominated by the correlation between $\dot{M}$ and the product of $f_{acc}$ and $R_*^2$.  However, by comparing the correlations from Table 10 and Table 11 we can see that the inclusion of $M_*$ and $P_{rot}$ does enhance the quality of the relationship when looking at eq. (3).  The expectation of a correlation between $\dot{M}$ and $R_*^2$ is obvious from eq. (7), but the lack of a correlation between $\dot{M}$ and $f_{acc}$ is not quite so clear.  Furthermore, if the good correlation seen in Figure 8 is driven primarily by the correlation between $\dot{M}$ and $R_*^2$, we would also expect this strong correlation to show up in the plots for eq. (1) (Figure 6).  As we have seen, however, there are no correlations present in the eq. (1) plots leading us to the conclusion that the $\dot{M}$--$R_*^2$ correlation is not the primary driver for the correlations for eq. (3).  

A further test of eq. (3) is given by \citetalias{JK02} as

\begin{equation}
R_*^{-1/2}L_{acc}^{1/2}\propto F_{acc}P_{rot}^{1/2}
\end{equation}

\noindent which can be obtained by substituting eq. (4) into eq. (3), solving eq. (5) for $f_{acc}$ and plugging the result into eq. (3).  Equation 8 provides the best test of the modified \citetalias{OS95} theory in the sense that the parameters involved are determined as independently as possible from one another.  Although both $L_{acc}$ and $F_{acc}$ are ultimately determined by model fits to the data, $F_{acc}$ is the quantity directly measured from the observations.  In addition, $L_{acc}$ and $F_{acc}$ are not well correlated ($r$=0.21, $P$=0.31) so any real correlation shown by eq. (8) is a result of the full comparison of all the combined variables.  Figure 10 shows a scatter plot of eq. (8).  While there is some evidence of a correlation (\textit{r}=.46), the probability of no correlation is high enough ($P$=0.02) to prevent any definitive statements about the support of eq. (3), rewritten here in the form of eq. (8).  However, when L7974 is removed (the lower right-most blue circle) from the analysis, the correlation improves to \textit{r}=0.65 with $P$=.006 suggesting a stronger relationship between the variables.

A similar test can be done for eq. (1).  By replacing $\dot{M}$ in eq. (1) with eq. (4) and dropping the constants, eq. (1) can be rewritten as

\begin{equation}
R_*^{5/2} \propto M_*^{1/3} L_{acc}^{1/2} P_{rot}^{7/6}.
\end{equation}

\noindent Figure 11 shows the results of plotting the SA and WA stars using eq. (9).  There is no correlation present at all, confirming the lack of support for the dipole models of eq. (1).  In fact, the correlation is worse than the correlation from eq. (1) (Figure 6) indicating that the correlation between $\dot{M}$ and $R_*^2$ is most likely bolstering the weak correlation found for eq. (1).

We have also tested the impact on the correlations of calculating individual values of $R_{Tr}$ for each star.  Each truncation radius was computed by assuming the star is locked to its disk.  In other words, $R_{Tr}$ is simply the radius at which the disk rotates with the same Keplerian rotation rate as the star.  The changes in $\dot{M}$ using the individual values of $R_{Tr}$ range from as little as 1\% to as much as 50\%.  However, when the analysis is repeated using the new values of $\dot{M}$, the correlations do not show any significant change for either eq. (1) or eq. (3) (eq. 1--$r$=0.33, $P$=0.10; eq. 3--$r$=0.90, $P$=7x10$^{-10}$).  This is not surprising since changing the truncation radius by even a factor of 10 will most often result in less than a factor of 2 change in the accretion rate.         

We believe that the moderate correlation shown by eq. (8) provides additional support for the modified version of the \citetalias{OS95} theory of magnetospheric accretion.  The very strong correlations shown in Fig. 8, however, provide the strongest support for the \citetalias{OS95} model even though we find that the eq. (3) correlations could possibly be driven by a correlation between $\dot{M}$ and $R_*^2$.  If the correlation did not improve with the inclusion of $M_*$, $P_{rot}$, and $f_{acc}$ we would hesitate to draw any firm conclusions concerning the validity of the modified \citetalias{OS95} theory.  However, since the correlations do improve and the result is a very tightly constrained relationship between the two sides of eq. (3), it appears the analysis of \citetalias{JK02} to include non--dipole surface magnetic field geometries within the context of the \citetalias{OS95} model is supported by the data in NGC 2264.

\section{DISCUSSION}

Using stellar and accretion parameters for TTSs in the Taurus--Auriga star forming region, \citetalias{JK02} also find support for the modified version of the \citetalias{OS95} theory through the same type of analysis we present here.  The correlation presented by \citetalias{JK02} for eq. (8) (their eq. 12) is stronger than the one we find for our sample.  The \citetalias{JK02} correlation appears to not be strongly influenced by a correlation between $R_*^2$ and $\dot{M}$, while the effect of such a relationship may have a larger influence on our results for eq. (3).  Nonetheless, both studies show much stronger support for the modified \citetalias{OS95} theory than they do for the models of \citet{kon91}, \citet{camcamp93}, and \citet{shu94}, all of which assume a purely dipolar stellar magnetic field.  As \citetalias{JK02} point out, this does not rule out the domination of the dipolar component at radii comparable to $R_{Tr}$ due to the faster decrease in field strength of higher--order components with increasing distance from the stellar surface.  In fact, S. G. Gregory et al. (2012, in press) show that the dipole component dominates at the truncation radius even if the field strength of the octopolar component is ten times that of the dipolar component at the magnetic pole.  As mentioned briefly in \S 2, there is mounting evidence that magnetic fields on TTSs are not purely dipolar \citep{valenti04,JK07,donati11a,donati11b,donati11c}.  While higher order magnetic field components are important in determining details of the accretion flow \citep{mohanty08,adams12}, estimates of the disk truncation radius can be made using the dipolar component only and ignoring the negligible contributions from the higher order components.   

Based on their recent measurements of magnetic fields for a small sample of TTSs \citep[e.g.][]{donati11c,donati11b,donati11a,donati10,donati08,donati07}, Donati et al. have speculated that the dipole component of the magnetic field should become weaker relative to the higher order field components as the star ceases to be fully convective and develops a radiative core.  This evolution of the magnetic field geometry is given as evidence that the magnetic fields of TTSs are dynamo--driven rather than primordial \citep{donati11c}.  If the strength of a star's dipolar magnetic field decreases, it is expected that the accretion disk would be able to move in to a smaller radius enabling the star to spin--up and decrease its rotation period.  Thus, according to the suggestion of Donati et al., one might expect the rotation periods of partially convective stars to be shorter as a group than the fully convective stars.  Eq. (1), however, shows that the relationship between $B_{dip}$ and $P_{rot}$ also depends on the star's mass, radius, and accretion parameters, i.e. suggesting that $P_{rot}$ should decrease as $B_{dip}$ decreases is an oversimplification of the problem.  Therefore, in order to test this hypothesis, we have plotted (Fig. 12, upper panel) the strength of the equatorial dipolar magnetic field component as derived by \citet{JK99} for the \citet{shu94} magnetospheric accretion model (effectively solving eq. (1) for $B_{dip}$ but with the appropriate constants to turn the proportionality into an equality) against the ratio of the mass of the radiative core to the total stellar mass.  This is the magnetic field required by disk-locking theory to maintain the observed rotation rate given the measured mass, radius, and accretion rate onto the star.  The core mass was calculated using the \citet{siess00} evolutionary models for the SA and WA stars, where $M_{core}$/$M_*$=0.0 is a fully convective star.  The NA stars are not included due to the lack of derived accretion rates required to estimate $B_{dip}$.  

It can be seen from the upper panel of Fig. 12 that the derived dipolar component of the magnetic field at the stellar equator does not correlate with the size of the star's radiative core.  We have also examined the predicted field strengths of the fully convective and partially convective samples using a Kolmogorov--Smirnov test.  The significance of the K--S test is $P_{KS}$=0.24, large enough to prevent the rejection of the null hypothesis that the two samples are drawn from the same distribution.  We also performed a K--S test on the rotation periods of the two samples, which are shown in the lower panel of Fig. 12.  In this case, $P_{KS}$=0.38, again providing little evidence that the samples are drawn from separate distributions.  The upper panel of Fig. 12 was made assuming two things: 1. the star--disk interaction is dominated by the dipolar magnetic field component, and 2. the star is locked into co--rotation with its disk, i.e. disk--locking.  Due to the lack of correlation seen in the upper panel of Fig. 12, one or both of these assumptions must be incorrect or the evolution of the magnetic field as suggested by Donati et al. is incomplete.  However, our data show good support for disk--locking and the domination of the star--disk interaction by the dipolar magenetic field component is well supported \citep[e.g.][]{adams12,donati11a,mohanty08}.  More measurements of the dipolar field component are needed for a larger sample of TTSs in order to reinforce or reject the recent findings of Donati et al.      

While our data (and that of \citetalias{JK02}) support a somewhat modified version of the \citetalias{OS95} model and therefore indirectly provide support for the magnetocentrifugally driven wind (X--wind) which is a component of that model \citep[see also][]{mohanty08}, other competing theories have been developed to explain angular momentum loss in low mass stars and thus the observed slow rotation in most CTTSs.  These models invoke an accretion driven \textit{stellar} wind as the main momentum regulation device \citep[e.g.][]{matt05,matt08} and argue, based on numerical calculations of magnetic torque interactions between the disk and star, that the torques associated with disk--locking theories are not sufficient to produce the observed slow rotation rates.  In fact, when only considering the magnetic coupling of the star and disk, these theories predict a \textit{positive} torque on the star \citep[e.g.][]{matt10}, causing it to spin up.  Using the same analysis discussed in \S$\;$5 we have tested the equilibrium equations of \citet{matt08} (Matt, private communication) involving $M_*$, $\dot{M}$, $R_*$, and $P_{rot}$.  For the case $R_{co}$$\approx$$R_{Tr}$, where $R_{co}$ is the corotation radius, we find weak evidence of a correlation ($r$=0.42, $P$=0.04).  The \citet{matt08} relationships assume a purely dipolar magnetic field.  Thus, based on the lack of support for eq. (1), it is not surprising that only weak correlations are present in our data for the \citet{matt08} relationships.  

We believe our analysis provides support for the \citetalias{OS95} model of magnetospheric accretion, extended by \citetalias{JK02}, and later \citet{mohanty08}, to include non--dipole field geometries.  These results also support the disk--locking scenario which seems to be well supported observationally based on period distributions of young stellar clusters \citep[e.g.][]{herbst02,lamm05}, although conflicting observations do exist \citep[e.g.][]{stass99,rebull02,mak04}.  However, \citet{rebull06}, \citet{cieza07}, and \citet{dahm2011} find the strongest evidence to date that circumstellar disks are directly involved in regulating the angular momentum of their central stars, effectively ruling out a mechanism that does not involve significant angular momentum exchange between the star and its disk.  Accretion--driven stellar winds could certainly provide a means of angular momentum removal for CTTSs but the significant correlations found here and by \citetalias{JK02} for eq. (3) provide observational support for the existence of disk--locked systems, a constraint that is not necessary in the stellar wind models. 

\section{SUMMARY AND CONCLUSIONS}

We have derived stellar and accretion parameters for 36 TTSs in NGC 2264 using spectrophotometric measurements taken in 2004 and 2005.  Our estimate for the age of the sample ($\sim$6.4 Myrs), calculated using the pre--main sequence evolutionary tracks of \citet{siess00}, is older than several more statistically significant age determinations which average to $\sim$3 Myrs.  This is due in large part to: 1. our small sample size, and 2. our use of the \citet{siess00} PMS models which tend to produce older ages compared with the commonly used tracks of \citet{baraffe98} and \citet{dantona94} \citep{rebull02,dahm05}.  The mass accretion rates for our stars are similar to estimates by \citetalias{rebull02}, though the agreement is worse at lower $\dot{M}$.  Using the derived parameters we test analytic predictions from the purely dipolar disk--locking models of \citet{kon91} and \citet{shu94} (eq. 1) and the non--dipolar field prediction of \citetalias{JK02}'s modified version of the \citet{OS95} model (eq. 3).  We find good support for the modified \citet{OS95} model of magnetospheric accretion and disk--locking, although the correlation is influenced to some degree by a strong relationship between $\dot{M}$ and $R_*^2$.  A lack of support for the dipolar theories, however, highlights the need for an extra constraint in the theory or the abandonment of disk--locking.  This constraint is provided by the inclusion of $f_{acc}$ in eq. (3).  

Although the support we find for the \citet{OS95} theory is not without uncertainty, our results confirm the findings of \citetalias{JK02} and provide more evidence for disk--locking than against it.  In addition, our results find no support for theories that assume a dipolar magnetic field geometry at the stellar surface.  Recent evidence has shown that this assumption is probably not valid for TTSs \citep[e.g.][]{daou06,mohanty08,donati08,donati11b}.  This scenario does not exclude the possibility of accretion powered stellar winds as the agent which removes angular momentum from TTSs \citep[e.g.][]{matt05,cranmer}.  In fact, both stellar winds and disk winds launched from near the truncation point probably play a role in the removal of angular momentum, as suggested by \citet{edwards06}. 

The primary assumption made in our initial analysis is that the magnetic field strength does not vary significantly from star to star.  We have attempted to roughly account for differing dipolar magnetic field strengths from star to star by assigning values based on the size of the star's radiative core as suggested by the recent results of Donati et al.; however, doing so actually weakened the correlations present in the data.  In order to make this analysis more robust, better statistics on the strength of dipolar magnetic field components in TTSs, especially those at the $\sim$3 Myr age, are needed.  Whether the constant field strength assumption is justified or not for stars at similar evolutionary stages may help explain the slope evolution observed in our data.  In addition, we do not find any relationship between the theoretical dipolar magnetic field component and radiative core mass in our sample.  Future studies of emission and absorption lines from TTSs environments will help place constraints on the launching region of the outflows, thus helping to confirm or reject the hypothesis of disk--locking, and subsequent removal of angular momentum by a disk wind from the truncation radius, supported in this work.  Until more observational evidence can be gathered, it appears that the non--dipolar magnetspheric accretion model of \citet{OS95} remains a strong candidate for explaining the observed relationship between stellar and accretion parameters of CTTSs at the $\sim$3 Myr evolutionary stage and, by extension, that the disk--locking scenario is taking place in young stars.

\appendix
\section{Spectra and Model Fits}

This appendix contains the flux--calibrated spectra and model fits for our entire sample (Fig. A1).  The SA stars are displayed first, followed by the WA and NA stars.  Note that model fits are not given for the NA stars.  The black solid line is the observed spectrum; the model fit to the spectrum is overplotted in red; the template star is plotted as a green solid line; the slab flux is plotted as a blue dashed line.    

\section{Fitting the Data}   
              
The best--fit models to our spectra are generated using a non--linear least squares fitting routine based on the Marquardt method \citep[see][]{beving92}.  We generate our fits using a total of 6 free parameters: \textit{n}, \textit{T}, \textit{l}, $\alpha$, $\beta$, and $A_V$ where $\alpha$ is a flux scale factor relating the slab emission to the template star emission and $\beta$ is a scale factor relating the combined slab and template model to the observed TTS in NGC2264.  These scale factors account for the stellar radius and distance (for the TTS and template) and for the filling factor of the accretion zones.  The scale factors have the following forms:

\begin{equation}
\alpha=
\frac{f_{acc}}{(1-f_{acc})}\left(\frac{R_T}{d_T}\right)^2
\end{equation}
\begin{equation}
\beta=
\left(\frac{R_*}{R_T}\right)^2\left(\frac{d_T}{d_*}\right)^2
\end{equation}

\noindent where $R_T$ is the radius of the template, $R_*$ is the radius of the star, $d_T$ is the distance to the template, and $d_*$ is the distance to the star.  The values $R_T$, $d_T$, and $d_*$ are known or assumed quantities.  Thus, the final values of $\alpha$ and $\beta$ yield the surface filling factor $f_{acc}$ and stellar radius $R_*$, respectively.  The template radii were taken from the Catalogue of Apparent Diameters and Absolute Radii of Stars (CADARS) \citep{fracassini}.  We also calculated the template radii using published $B$--$V$ colors, parallaxes, bolometric corrections based on spectral type and effective temperatures taken from \citet{cohen79} (K0--K6), \citet{bessell} (K7--M1), and \citet{wilking99} (M2--M7).  All of the templates are MS stars within 30 pc.  Typical uncertainties in $T_{eff}$ are $\pm$150 K; for the bolometric corrections used, $\Delta$BC=$\pm$$^{m}_{^.}$05.  For typical photometric uncertainties of 5\% and errors of 5\% in the parallax measurements, the uncertainties in template radii calculations are $\sim$15\%.  The uncertainty in the distance to GJ596A is much higher, at $\sim$25\%, which translates into an uncertainty of $\sim$50\% in the stellar radius.  By comparing the observed \textit{B--V} colors of the templates with calibrated \textit{B--V} MS colors from \citet{allen}, we find that reddening is negligible to all of the templates.  In most cases our calculated radii agree with the CADARS radii to within 5\%.  In the cases of HD45088, GL394, and GJ596A the discrepency is closer to 15\%--30\%.  For consistency, we chose to adopt the CADARS values for these templates.  Table 3 lists the template stars and their parameters.  

The distance to NGC 2264 has not been precisely determined.  Current published values range from 750 pc \citep{mayne08} to 950 pc \citep{flaccomio} with a median value of $\sim$800 pc.  The most recent determination is 910 $\pm$ 110 pc by \citet{baxter}.  Here we choose to adopt a distance of $d_{NGC 2264}$ = 760 pc.  We have varied the distance used in our models from 700 pc to 900 pc and found deviations of $<$ 10\% in the values of the stellar accretion rates, filling factors, and radii that result.  In addition, because the distance to each individual star does not vary, the chosen distance to the cluster has no effect on the final parameter correlations and thus does not affect the main results of this study.

The overall reddening to NGC 2264 has been estimated by multiple studies \citep[e.g.][]{walker56,sung97,rebull02}.  The first photometric determination of the cluster reddening was done by \citet{walker56} using main sequence O and B stars and was found to be E($B-V$)$\sim$0.$^{m}$08.  \citet{young78}, using the standard extinction law R=3.1, determined that the cluster was differentially reddened and that dust clouds within the cluster varied with position, causing the reddening to change moderately from region to region.  This was interpreted as the result of intense radiation from higher mass stars clearing out the space around them and creating regions of higher and lower dust extinction.  Indeed, \citet{young78} identifies the excessively reddened stars in his sample as those that lie behind known dusty regions of the cluster.  However, \citet{rebull02}, using a much larger sample of stars, find no evidence of spatially dependent reddening within the cluster.  This may be due in part to their use of $R-I$ excesses which are less sensitive to foreground absorbers.  

Reddening values to individual stars can be affected by the local environment of the object, causing specific values of $A_V$ to differ significantly from one star to the next.  This individual reddening is what we attempt to estimate using our models.  \citet{sung97} confirm the differential nature of the reddening across the cluster found by \citet{young78}.  \citet{sung97} also determine E($B-V$) to be 0.$^{m}$071$\pm$.033, in agreement with earlier estimates.  In the more recent study of \citet{rebull02}, the cluster reddening is found to have a slightly higher value of E($B-V$)=0.146$\pm$0.03, corresponding to $A_V$=0.41.  These estimates, however, are ``most likely'' values and cannot be used in the context of individual stars.  Here, we solve for individual model estimates of \textit{A$_V$}.

Strong emission lines from elements other than hydrogen can affect the fits to the data.  For this reason we ignore $\sim$10 \AA$\;$of spectrum on either side of any strong emission lines.  The most common examples are the Ca{\sc ii} H \& K lines at 3933 and 3969 \AA.  The same procedure is applied to the Balmer lines discussed in \S 4.1.  Ignoring these lines provides better fits to the stellar continuum at red wavelengths and the excess continuum at bluer wavelengths. 

Based on similar work by \citet{valenti93} and \citet{phart91}, we adopt initial values for our slab parameters of \textit{n}=$10^{14}$ cm$^{-3}$, \textit{l}=$10^7$ cm, and \textit{T}=9000 K.  In a few cases, the fitting routine had trouble establishing a good fit using our standard intial parameters.  For these stars, the initial values were adjusted manually until a suitable location in parameter space was found.  It should be noted that our criteria for a 'good fit' are mainly subjective, i.e. the fits to the data are examined manually.  We look for two things: 1. Is the Balmer jump well matched by the slab?; and 2. Are the underlying photospheric features fit well by the chosen template?  If these criteria are met and the model approaches the same reasonable reduced $\chi^2$ value for multiple models with different initial parameter values (this is usually an indication that the fitting routine has found a minimum $\chi^2$ in the vicinity of the absolute minimum of the parameter space), the fit is considered 'good'.  A discussion of the uniqueness of our models is given below.  Typical reduced $\chi^2$ values for our models are $\sim$1.0-2.0.  In order to produce superior fits, it was common to use a template that differed in spectral type by 1-2 subclasses from the determined spectral type of the star being modeled (see Table 2).  This aspect of the model is discussed more thoroughly below.

Initial values for $\beta$ and $A_V$ were determined by fitting the CTTS with a scaled template star in the absence of a slab.  Due to the relative weakness of the excess spectrum compared to stellar photospheric emission at wavelengths $>$ 4500 \AA$\;$(the exception being stars experiencing large amounts of mass accretion), this procedure generally established good fits to the red portion of the spectrum and provided stable starting points for the $\chi^2$ routine.  Typical starting values for $\beta$ were $10^{-4}$, corresponding to $R_*$$\sim$1.5 R$_\odot$ for $R_T$$\sim$.75 R$_\odot$ and $d_T$$\sim$12 pc.  Starting values for $\alpha$ were determined by assuming $f_{acc}$=.05.  Again, some cases required manual adjustment of the initial values of the scale factors.  

The model slab flux is then generated with the best--fit parameters.  The template photosphere is then subjected to the optical depth of the slab:

\begin{equation}
F^{'}_T = 
[f_{acc}e^{-\tau}+(1-f_{acc})]F_T.  
\end{equation}

\noindent The slab is then scaled and added to the adjusted photosphere.  The combined spectrum is then subjected to the reddening determined by the model and scaled to match the observed spectrum.  The modeled flux can be summarized by the following equation:

\begin{equation}
F_{*}=\beta\left[\alpha F_{slab}+F^{'}_{T}\right]
\end{equation}

\noindent where $F_*$ is the total (stellar plus accretion) flux, $F_{slab}$ is the slab flux, and $F^{'}_{T}$ is the adjusted template star flux from eq. (B3).  When constructing our models, we ignore any inclination or geometric effects concerning the location of the accretion column on the star.  The slab is taken to be in front of the stellar photosphere along the line of sight to the observer.  As a result, the filling factor we determine is the projected filling factor on the visible surface of the star.  Ignoring these effects introduces some uncertainty in the final models.  However, when averaged over the entire sample, assuming a uniform distribution of inclinations and number of accretion columns per star, this uncertainty should not affect the final correlations.

We cannot ever be sure if our $\chi^2$ minimization routine approaches the absolute minimum $\chi^2$ value for each fit.  Instead, we can be confident in our final model parameters if, given a range of reasonable starting values, the fitting routine approaches a model of similar quality for each set of initial values.  The uniqueness of the final stellar and accretion parameters is tested in the same manner.  Figures B1 and B2 show this process graphically for the accreting star L1316.  In these plots, all other intial parameters are fixed at the standard values described earlier in \S 4.1.        
Tables B1 and B2 list the initial parameters and final parameters for each model calculation.  It is obvious that while the exact values of the slab parameters certainly vary with the values chosen for the intial parameters, the final stellar and accretion parameters do not change significantly for fits with similar $\chi^2$ values.  Fits with the same $\chi^2$ value are practically indistintinguishable from one another.  It is important to note that even though the final values for \textit{T} and \textit{n} vary by as much as a factor of 2 and as much as a factor of 10 for \textit{$\beta$} and a factor of 5 for \textit{A$_V$}, \textit{the optical depth through the slab, $R_*$, and $\dot{M}$ do not vary appreciably at all}.  This shows that although the values of the slab parameters and scale factors are not unique to a specific $\chi^2$ value, the final stellar and accretion parameters are well constrained by the fitting procedure.

We also investigated the effect of template spectral type on the final model parameters by fitting each spectrum with multiple template stars.  The best fitting template is chosen manually by examining how well the underlying photospheric features are matched.  It is sometimes the case that two different templates will yield similar stellar parameters and $\chi^2$ values but one template does not appear to match the absoprtion features of the underlying star as well as the other.  In these cases the template that visually shows the best match to the photospheric features is chosen for the model.  More typically, however, changes in spectral type of the template by 1--2 subclasses result in a 10--30\% change in the $\chi^2$ value for the fit.  The exceptions to this procedure are L5924 and L4547.  Due to our lack of a G--type template we modeled these stars using our earliest template, HD10476, a K1 star.  Attempts to model these stars with later spectral types resulted in increasingly worse fits.  Surprisingly, fits to L5924 and L4547 using the K1 template produce decent fits without the inclusion of a slab.  For this reason, as well as the fact that these stars do not show signs of significant mass accretion, they are grouped as NA stars and are not included in the disk--locking theory investigations discussed in \S 5.  Thus, uncertainties due to the template spectral type, which for these two stars would be large, do not factor into our final analysis.  

The best--matching template for a star is usually quite obvious.  The stars for which it is not so clear are those with higher accretion rates, and as a result more highly veiled spectra.  However, these tend to be less sensitive to changes in template spectral type due to the dominance of the accretion emission.  The stellar parameters for these stars tend to be more uncertain than for the stars with well--fit photospheres.  Based on multiple fits to each star with different template spectra, we estimate that the model spectral types used are accurate to within 2 subtypes.  From Table 2 it is obvious that the spectral types determined by \citet{rebull02} and the spectral types determined from our model fits do not always agree.  In fact, 30\% of our best--fit spectral types differ by more than 2 subtypes from the \citet{rebull02} spectral types.  We do not include L6175 in this number due to our lack of template beyond spectral type M3.5.  \citet{rebull02} estimate the uncertainties in their derived spectral types as 2 subtypes for K stars, and less than 1 subclass for M stars.  We attribute the differences in our model--determined spectral types and those of \citet{rebull02} to the fact that we are fitting a bluer portion of the stellar spectrum, whereas \citet{rebull02} used the region 5000$\;$\AA $<$ $\lambda$ $<$ 9300 \AA$\;$to classify their stars.  This spectral region is less affected by accretion emission and is particularly good for classifying M stars.  We also do not have templates for each K and M subtype and so must use the closest available spectral type.  Veiling due to accretion emission makes our spectra less sensitive to changes in template spectral type.  We conclude that the differences observed here are not problematic to our analysis nor are they unexpected based on differences in the considered wavelength region.
   
\acknowledgements

We wish to thank the McDonald Observatory staff in West Texas and the staff of Kitt Peak National Observatory in Arizona for their hospitality and help during the observing runs for this research.  We are also grateful to the anonymous referee for their help in improving this manuscript.  This research has made use of the Simbad Astronomical database and the NASA Astrophysics Data System.  We wish to acknowledge partial support for this research from the NASA Origins of Solar Systems program through grant numbers NNX08AH86G and NNX10AI53G made to Rice University.

\begin{figure}
  \begin{center}
  \figurenum{1}
  \includegraphics[scale=.70]{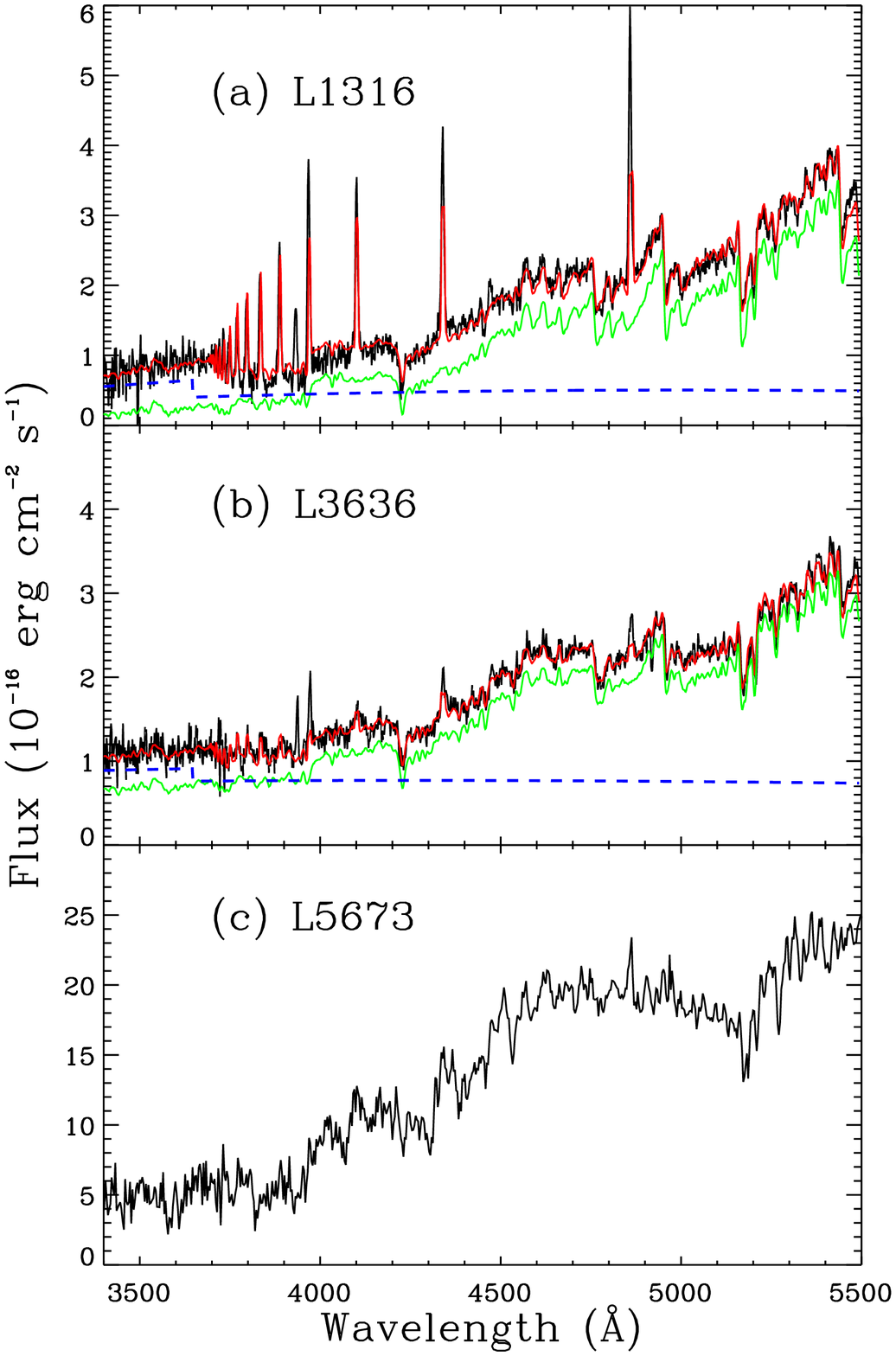}
  \caption{Example spectra and their associated models.  Panel (a) shows L1316, a typical SA star; panel (b) shows L3636, a typical WA star; panel (c) displays L5673, a typical NA star.  The black lines are the observed spectrum; the final model spectrum is overplotted in red; the underlying template star spectrum is overplotted in green and the slab spectrum (with the Balmer lines excluded for clarity) is overplotted with a blue dashed line.  Models are not produced for the NA stars.  Spectra and model fits for the entire sample are given in Appendix A.}
  \end{center}
\end{figure}
\clearpage
\begin{figure}[hbt]
\centering
\figurenum{2}
\includegraphics[scale=.80]{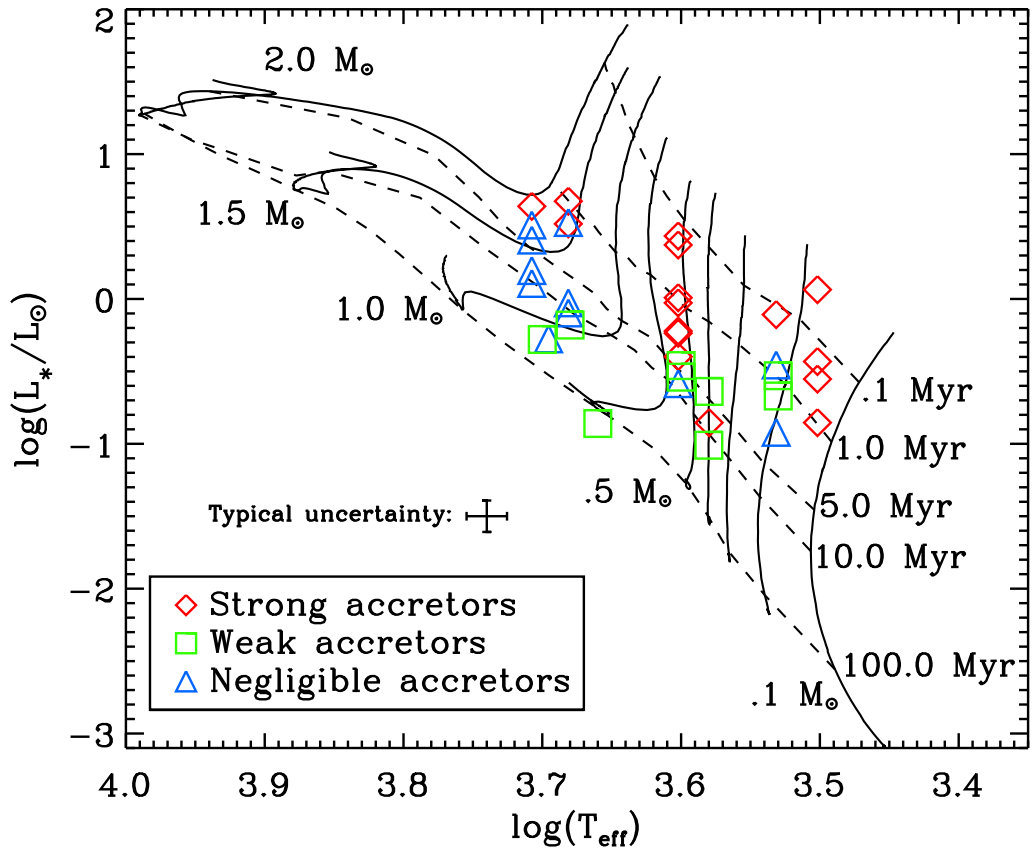}            
\caption{HR diagram of our sample using the \citet{siess00} PMS evolutionary models.  The accreting sample is younger on average than the rest of the sample, as is to be expected if these stars are undergoing more significant mass accretion.  The average age of our sample is $\sim$6.4 Myrs, slightly older than the literature average for NGC 2264 of $\sim$3 Myrs.  Estimates of the uncertainties in $T_{eff}$ and $L_*$ are shown above the legend.}
\end{figure}
\clearpage
\begin{figure}[hbt]
\figurenum{3}
\begin{center}
\includegraphics[width=6.5in]{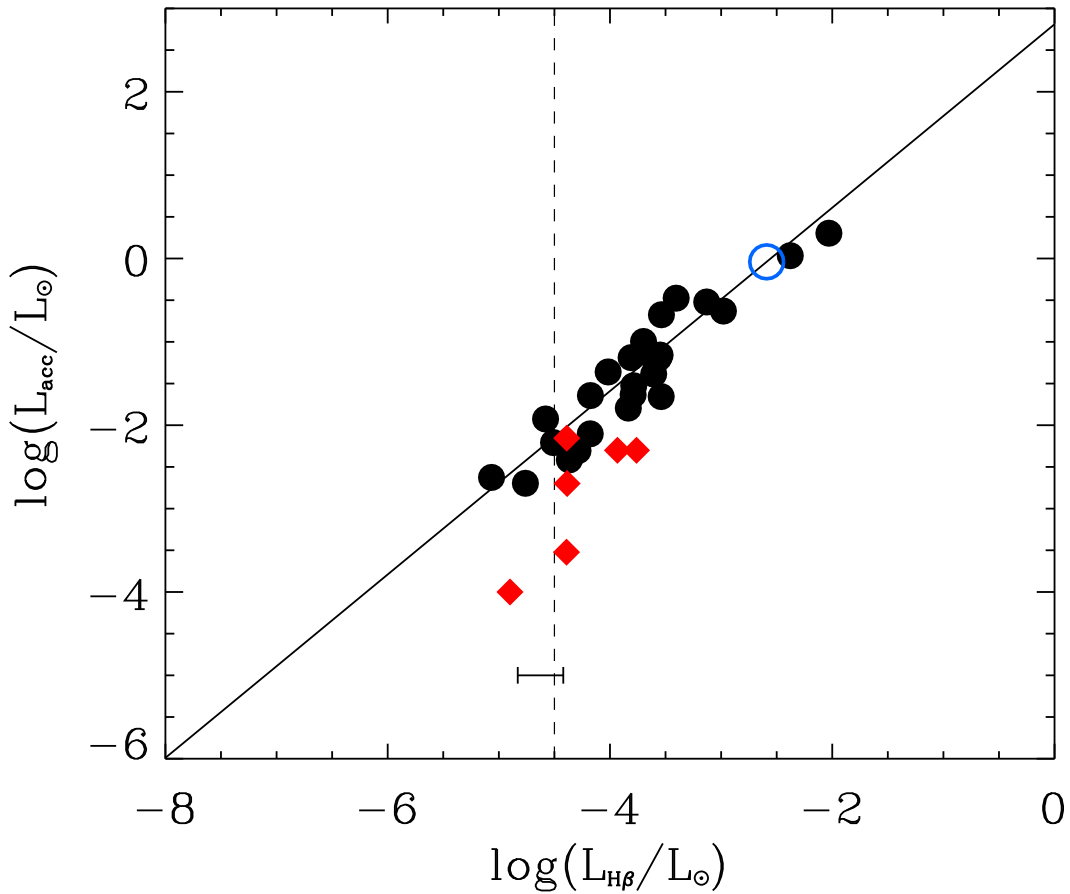}            
\caption{The H$\beta$ line luminosities of the SA and WA stars (black circles) plotted versus accretion luminosity.  The best--fit line (solid) to the SA and WA stars is overplotted.  The NA stars with measureable H$\beta$ emission are plotted (red diamonds), though they are not included in the determination of eq. (6).  The NA star L6172 is plotted (open blue circle) using the values from \citetalias{rebull02} to determine an accretion luminosity.  The vertical dashed line is a lower--bound on the amount of H$\beta$ luminosity that can be attributed to accretion (see \S$\;$4.5).  The bar at the bottom of the plot represents the range of H$\beta$ line luminosity spanned by the WTTSs used to determine the average chromospheric contribution to the H$\beta$ line flux.  The equation of the best--fit line is almost identical to the relationship found by \citet{fang09}.}
\end{center}
\end{figure}
\clearpage
\begin{figure}[hbt]
\figurenum{4}
\begin{center}
\includegraphics[width=6.5in]{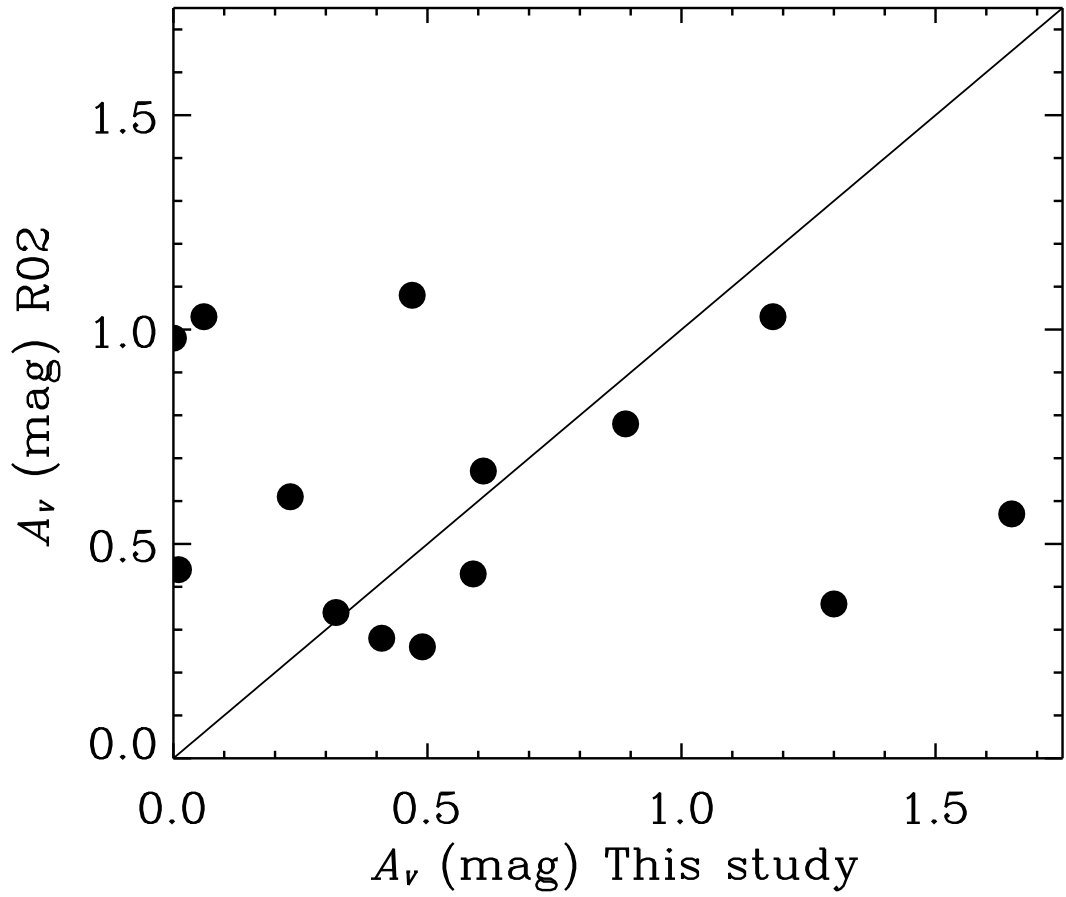}            
\caption{A comparison of $A_V$ values determined from \citetalias{rebull02} and those from this study.  The line of equal values is overplotted with a solid line.  There is poor agreement for $\sim$60\% of the stars.}
\end{center}
\end{figure}
\clearpage
\begin{figure}[hbt]
\figurenum{5}
\begin{center}
\includegraphics[width=6.5in]{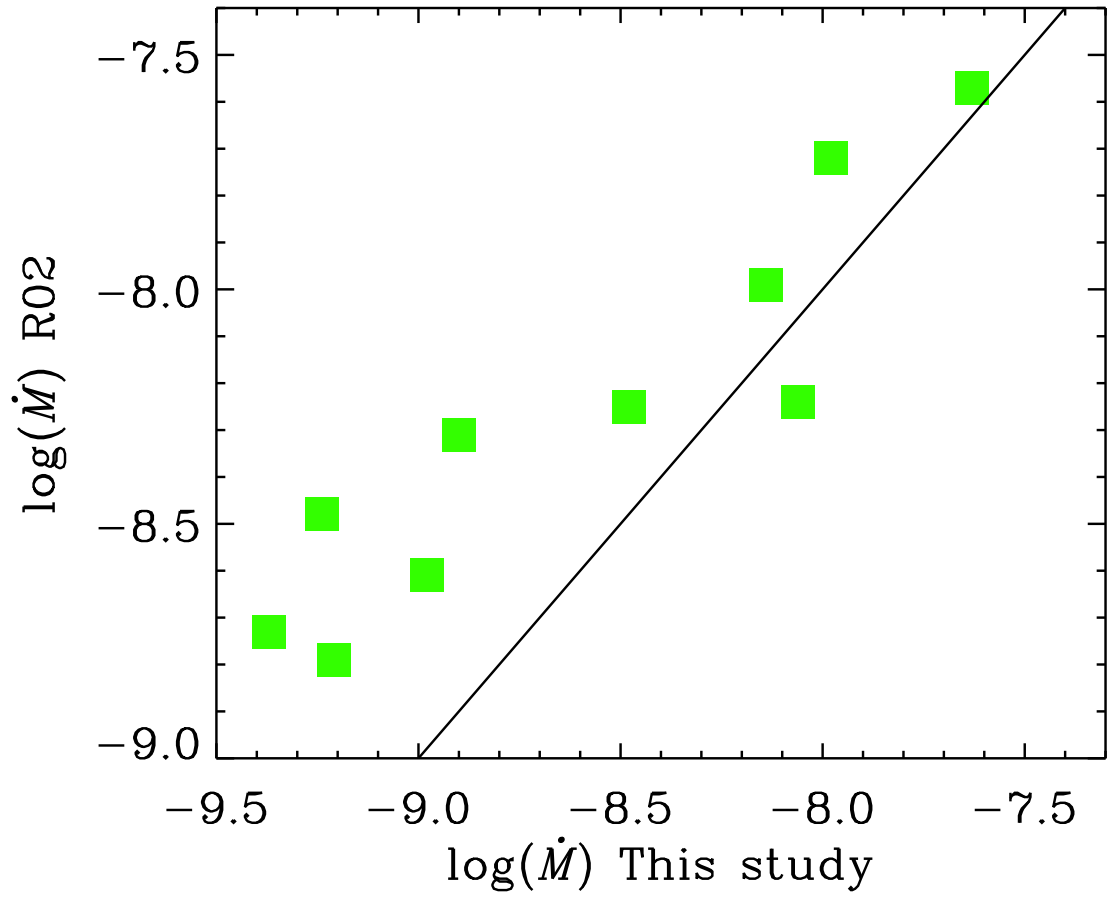}            
\caption{A comparison of $\dot{M}$ values determined from \citetalias{rebull02} and those from this study.  The line of equal values is overplotted with a solid line.  The \citetalias{rebull02} accretion rates are systematically higher by an average factor of $\sim$2.5.  This factor tends to be higher for stars with lower accretion rates in our sample.}
\end{center}
\end{figure}
\clearpage
\begin{figure}[hbt]
\figurenum{6a}
\begin{center}
\includegraphics[scale=.75]{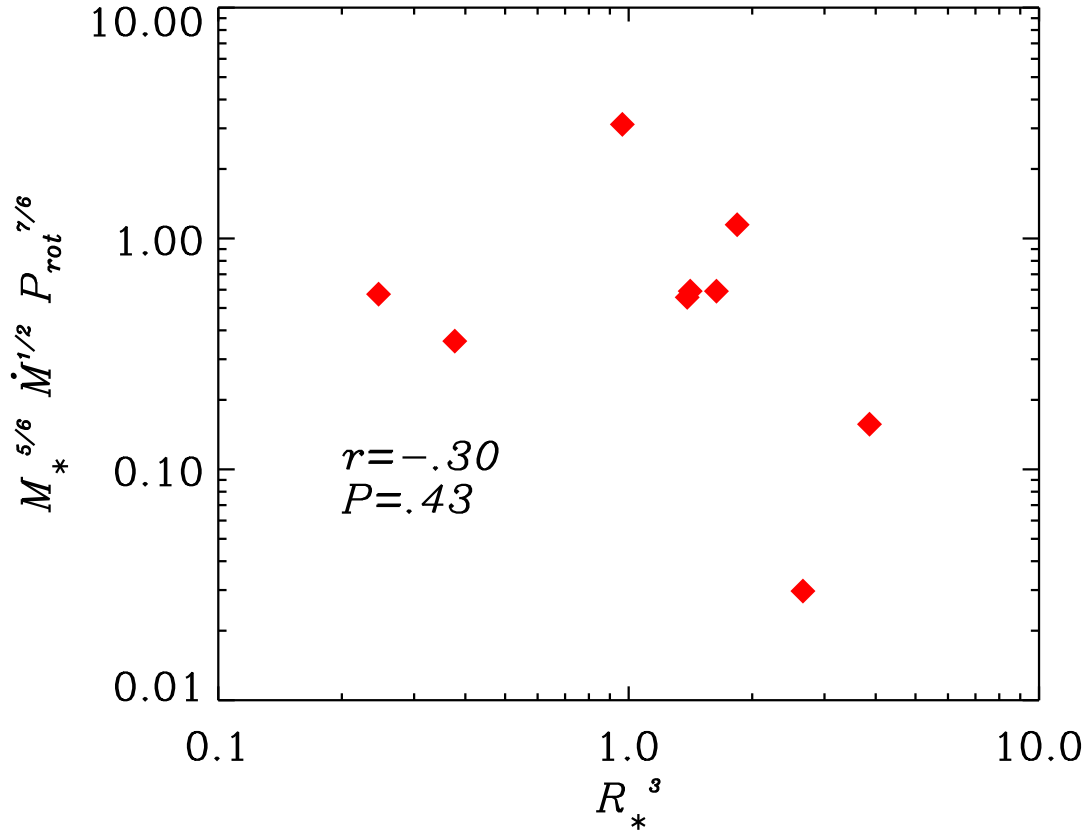}            
\caption{Plot of eq. (1), predicted by the models of \citet{kon91} and \citet{shu94}, for the weakly accreting stars assuming a constant magnetic field from star to star.  In Figures 6, 7, and 8 the radius is in units of R$_\odot$, $M_*$ is in units of M$_\odot$, $P_{rot}$ is in days, and $\dot{M}$ in units of 10$^{-7}$ M$_\odot$$\;$yr$^{-1}$.  No significant correlation is present.}
\end{center}
\end{figure}
\clearpage
\begin{figure}[hbt]
\figurenum{6b}
\begin{center}
\includegraphics[scale=.75]{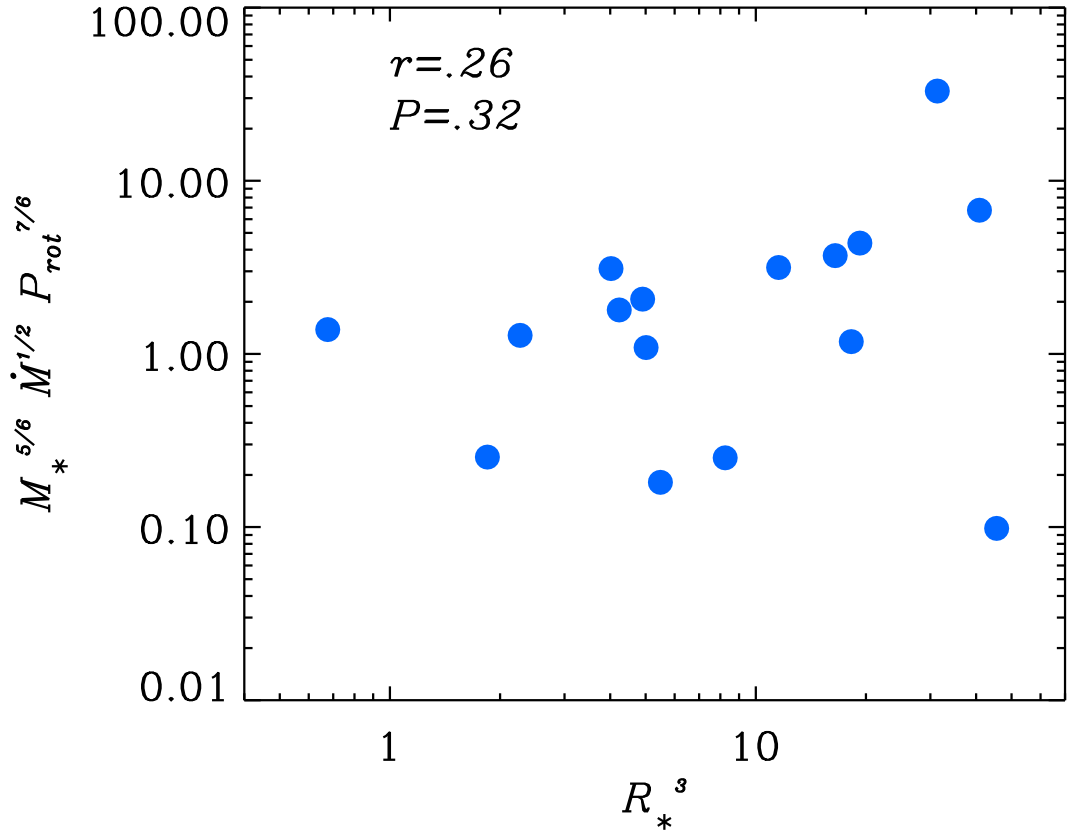}            
\caption{Same as Fig. 6a but for the strongly accreting stars.  The correlation is very weak.}
\end{center}
\end{figure}
\clearpage
\begin{figure}[hbt]
\figurenum{6c}
\begin{center}
\includegraphics[scale=.75]{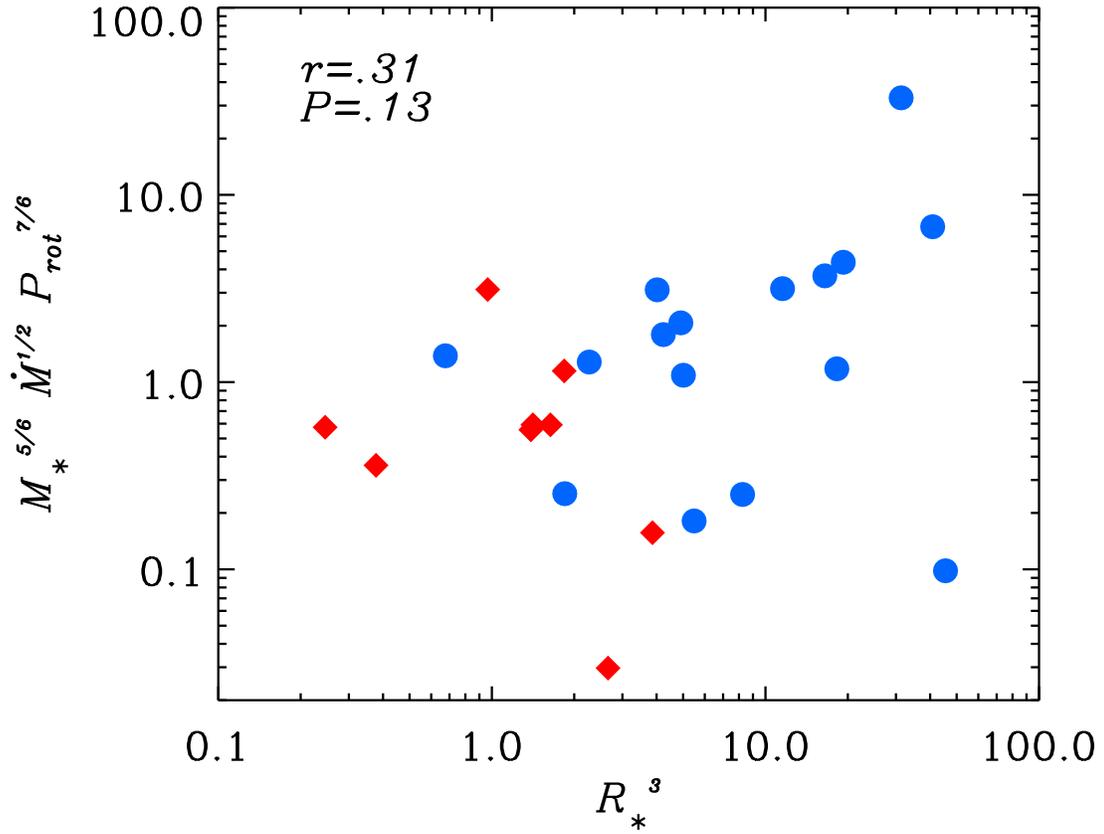}            
\caption{Same as figures 6a and 6b but for the combination of SA (blue circles) and WA (red diamonds) stars.  Again, no significant correlation is found.}
\end{center}
\end{figure}
\clearpage
\begin{figure}[hbt]
\figurenum{7}
\begin{center}
\includegraphics[scale=.75]{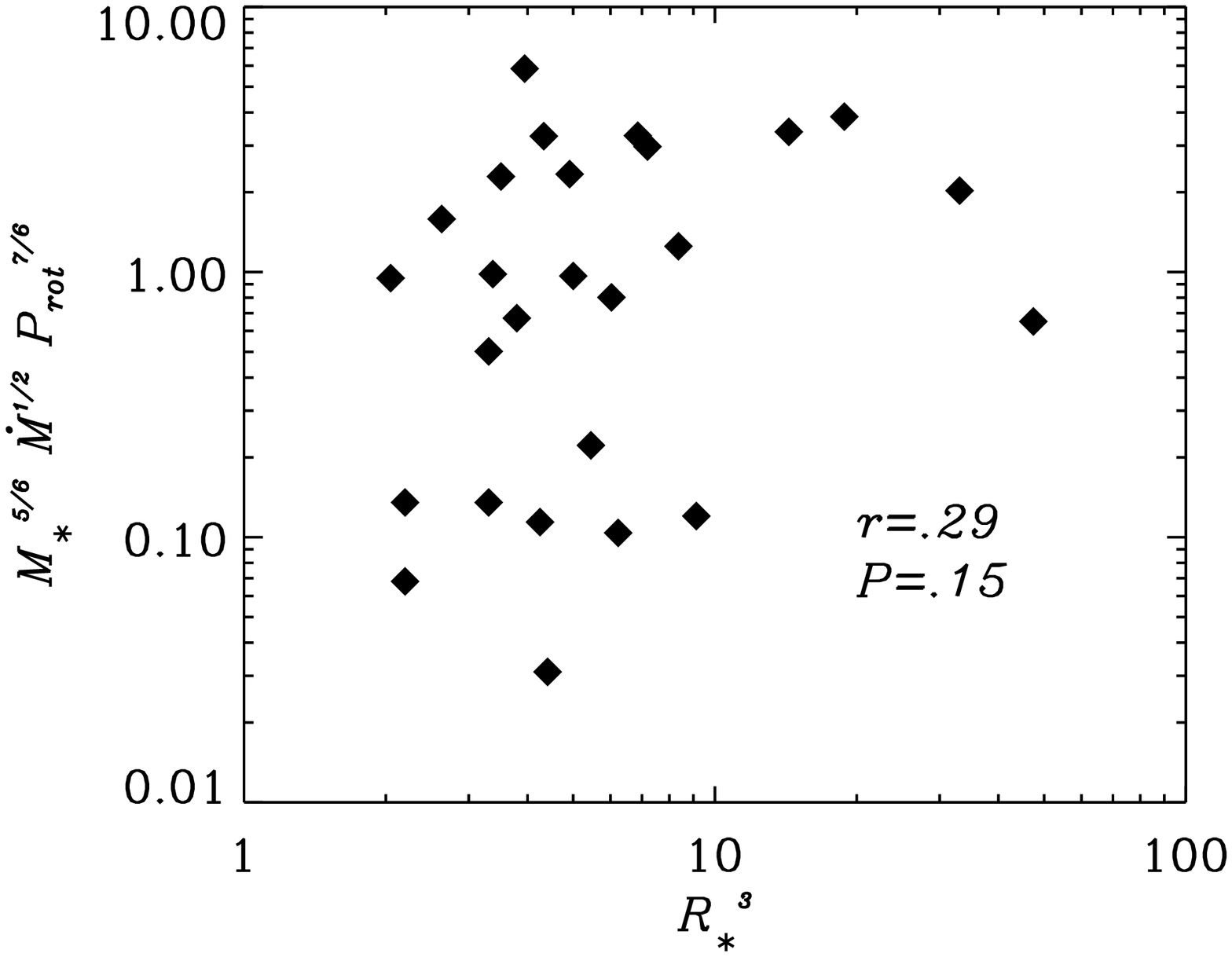}            
\caption{Plot of eq. (1) for 26 stars from \citetalias{rebull02}.  The rotation periods are taken from \citet{lamm04}.  There is no correlation present.}
\end{center}
\end{figure}
\clearpage
\begin{figure}[hbt]
\figurenum{8a}
\begin{center}
\includegraphics[scale=.75]{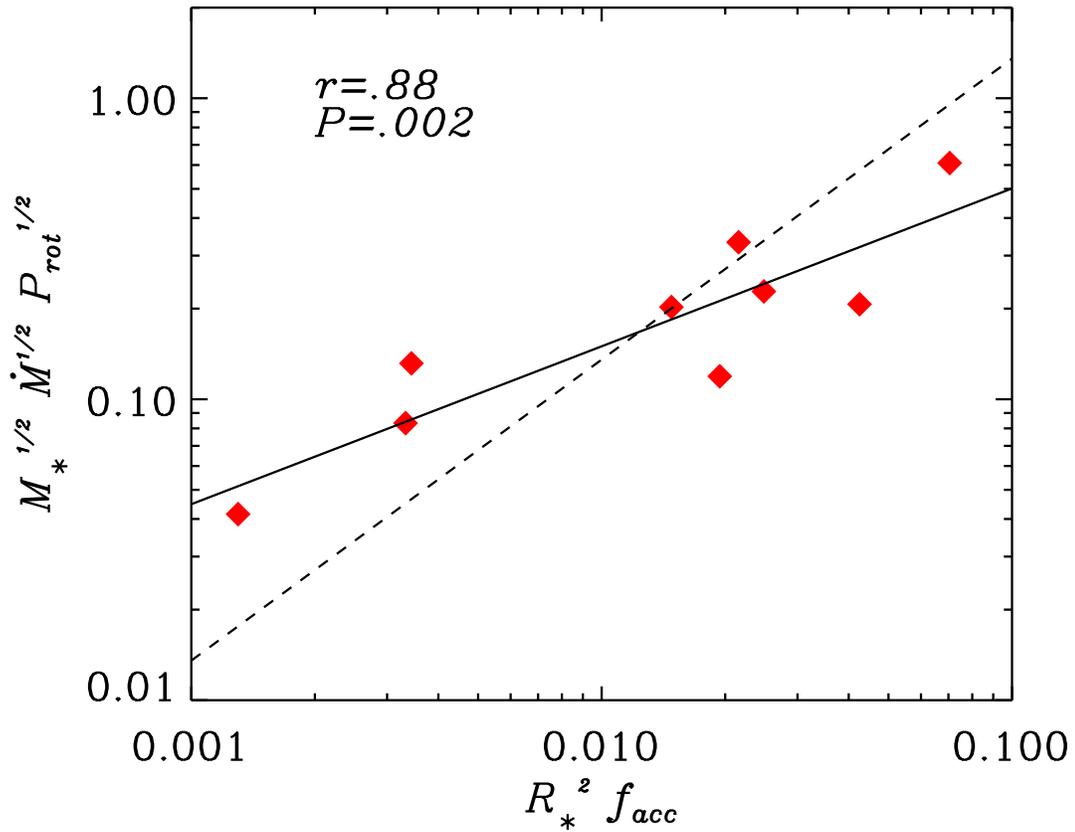}            
\caption{Plot of eq. (3) for the weak accretors.  The correlations for eq. (3) are much stronger than for eq. (1), with much smaller values of $P$.  Overplotted is the line of best--fit (solid line) and the line of best--fit with the slope fixed at 1.0 (dashed line), as predicted by the models.}
\end{center}
\end{figure}
\clearpage
\begin{figure}[hbt]
\figurenum{8b}
\begin{center}
\includegraphics[scale=.75]{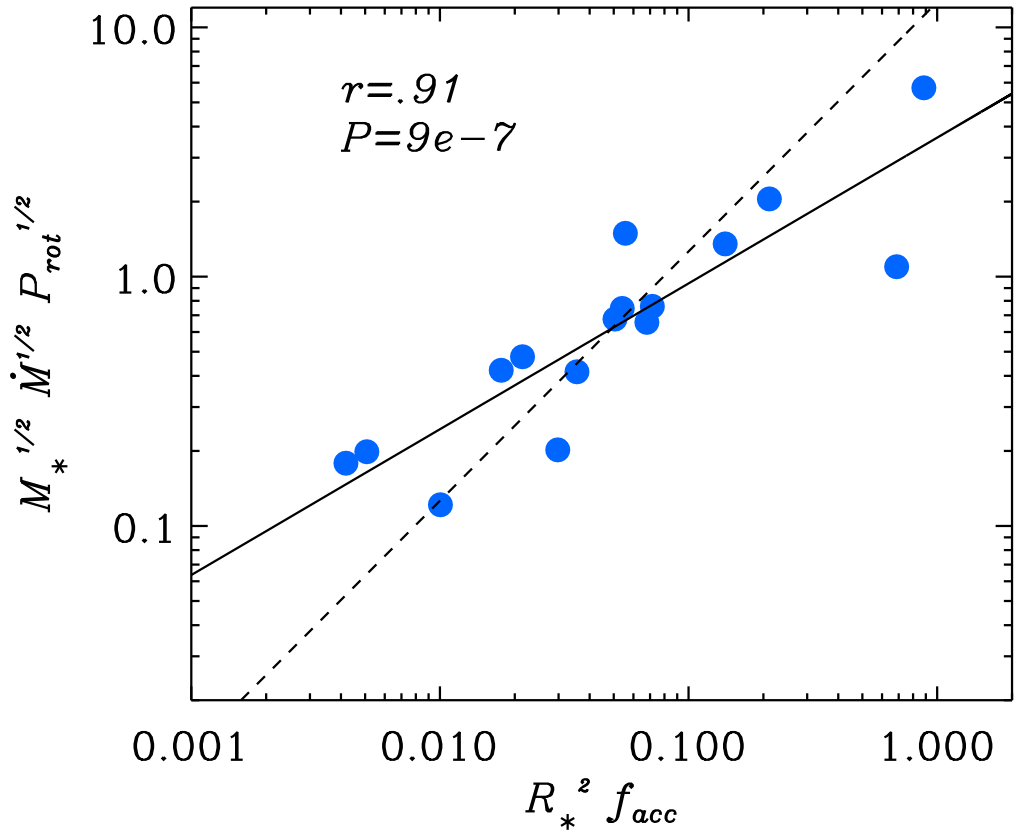}            
\caption{Same as Fig. 8a but for the strongly accreting stars.  The correlation is strong with a very low value of $P$, suggesting support for the modified \citet{OS95} theory.}
\end{center}
\end{figure}
\clearpage
\begin{figure}[hbt]
\figurenum{8c}
\begin{center}
\includegraphics[scale=.75]{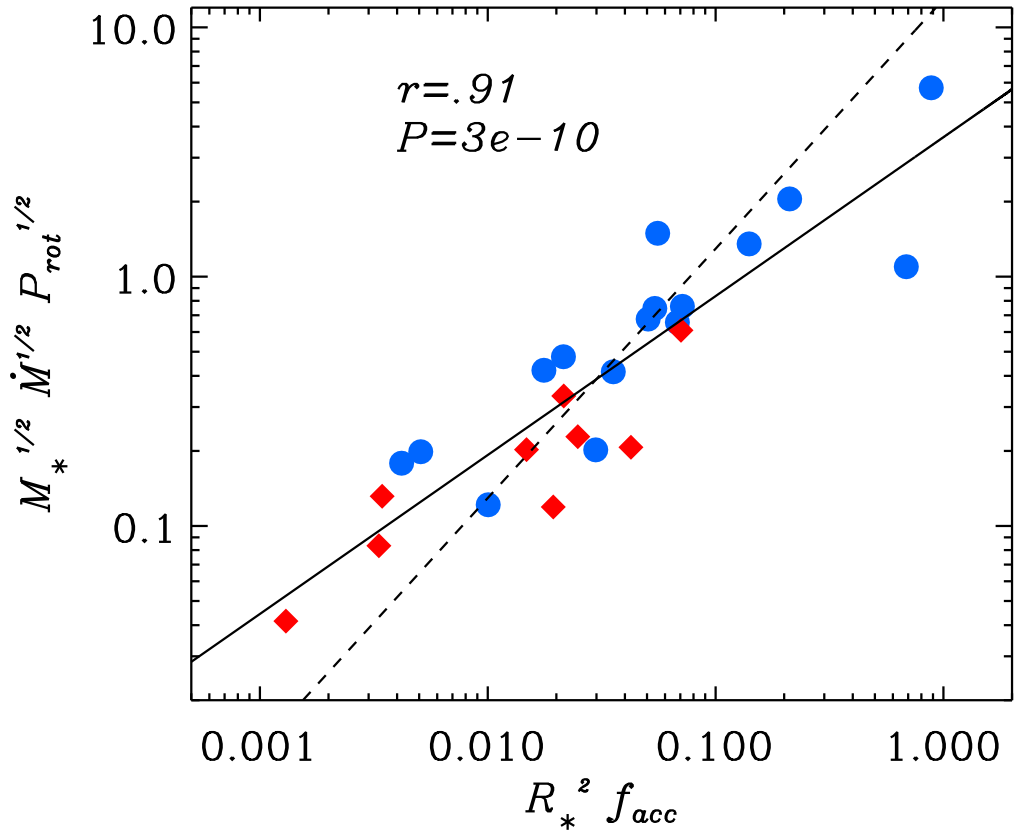}            
\caption{Same as figures 8a and 8b but for the combined group of SA (blue circles) and WA (red diamonds) stars.  This plot shows the strongest correlation with the best correlation significance.}
\end{center}
\end{figure}
\clearpage
\begin{figure}[hbt]
\figurenum{9}
\begin{center}
\includegraphics[scale=.80]{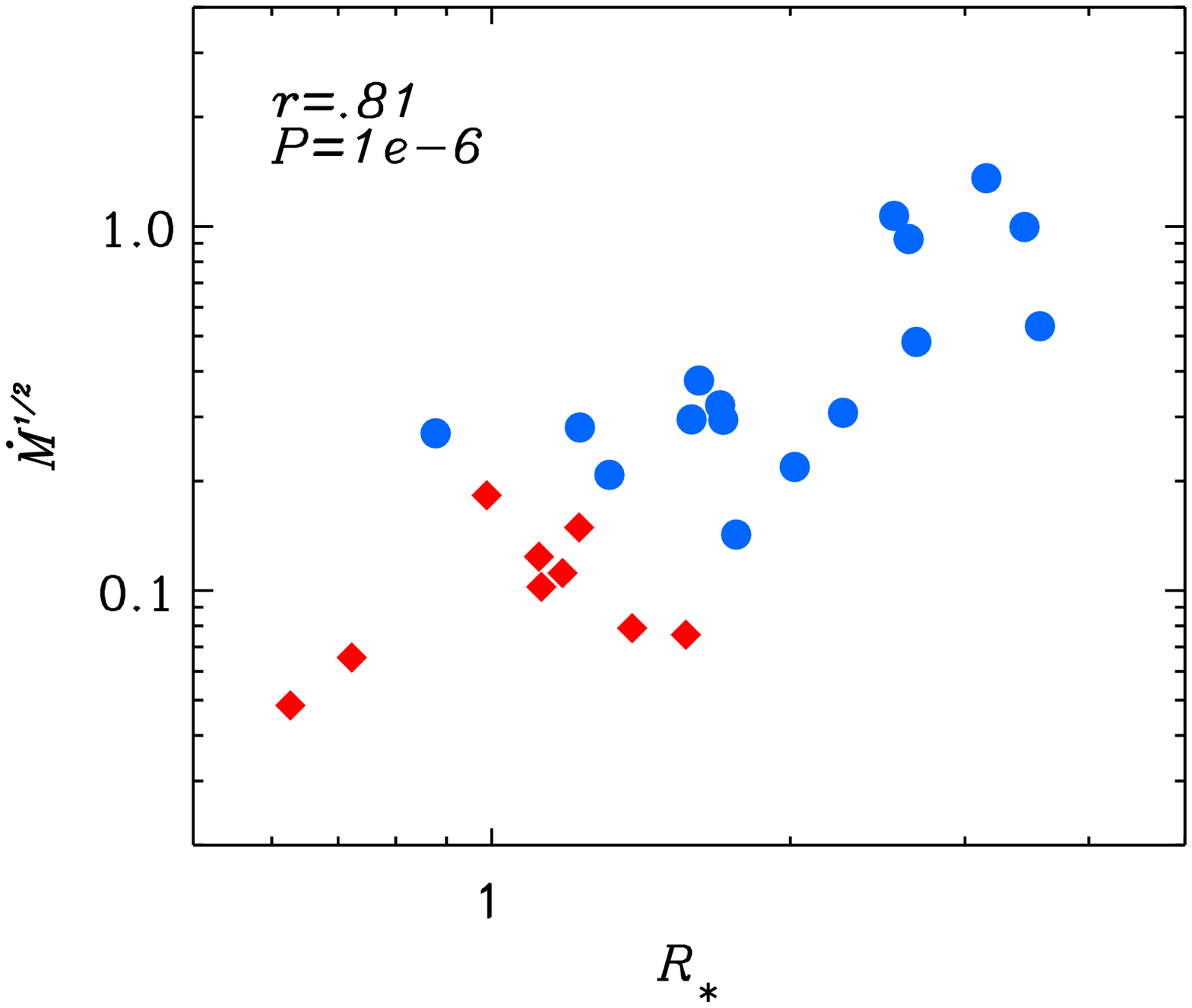}            
\caption{A strong correlation is seen between $\dot{M}$ and $R_*^2$, casting doubt on the significance of the correlation seen in Figure 8c.  The associated correlation coefficient and correlation significance are given in Table 12.  Plot symbols are the same as figs. 6 and 8.}
\end{center}
\end{figure}
\clearpage
\begin{figure}[hbt]
\figurenum{10}
\begin{center}
\includegraphics[width=6.5in]{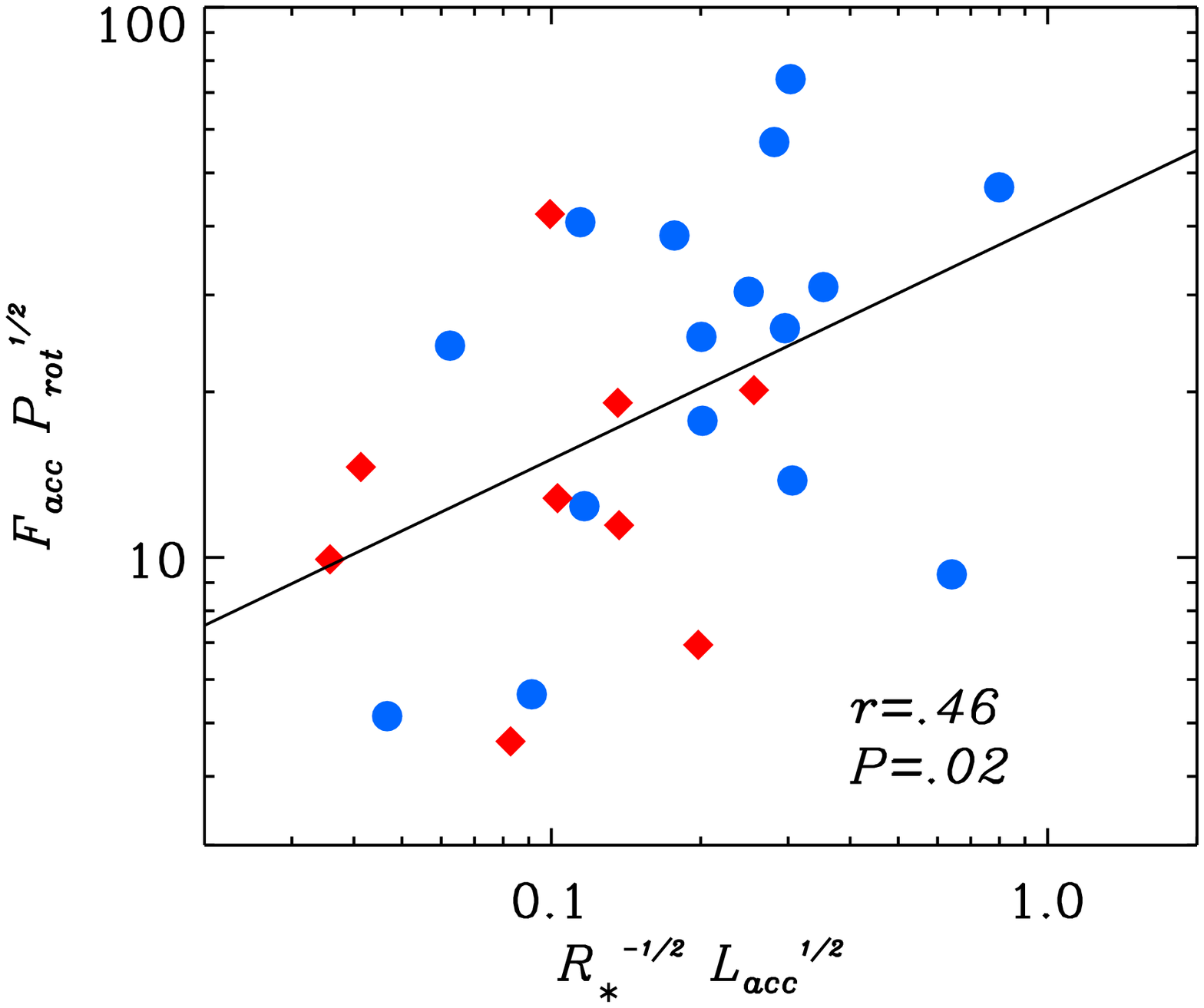}            
\caption{Equation 11 plotted using the combination of the SA (blue circles) and WA (red diamonds) stars.  A moderate correlation is found, although the correlation is increased significantly (\textit{r}=.57, $P$=.004) if L7974, the lower right--most circle, is excluded from the analysis.  The line of best--fit is overplotted.}
\end{center}
\end{figure}
\clearpage
\begin{figure}[hbt]
\figurenum{11}
\begin{center}
\includegraphics[scale=.80]{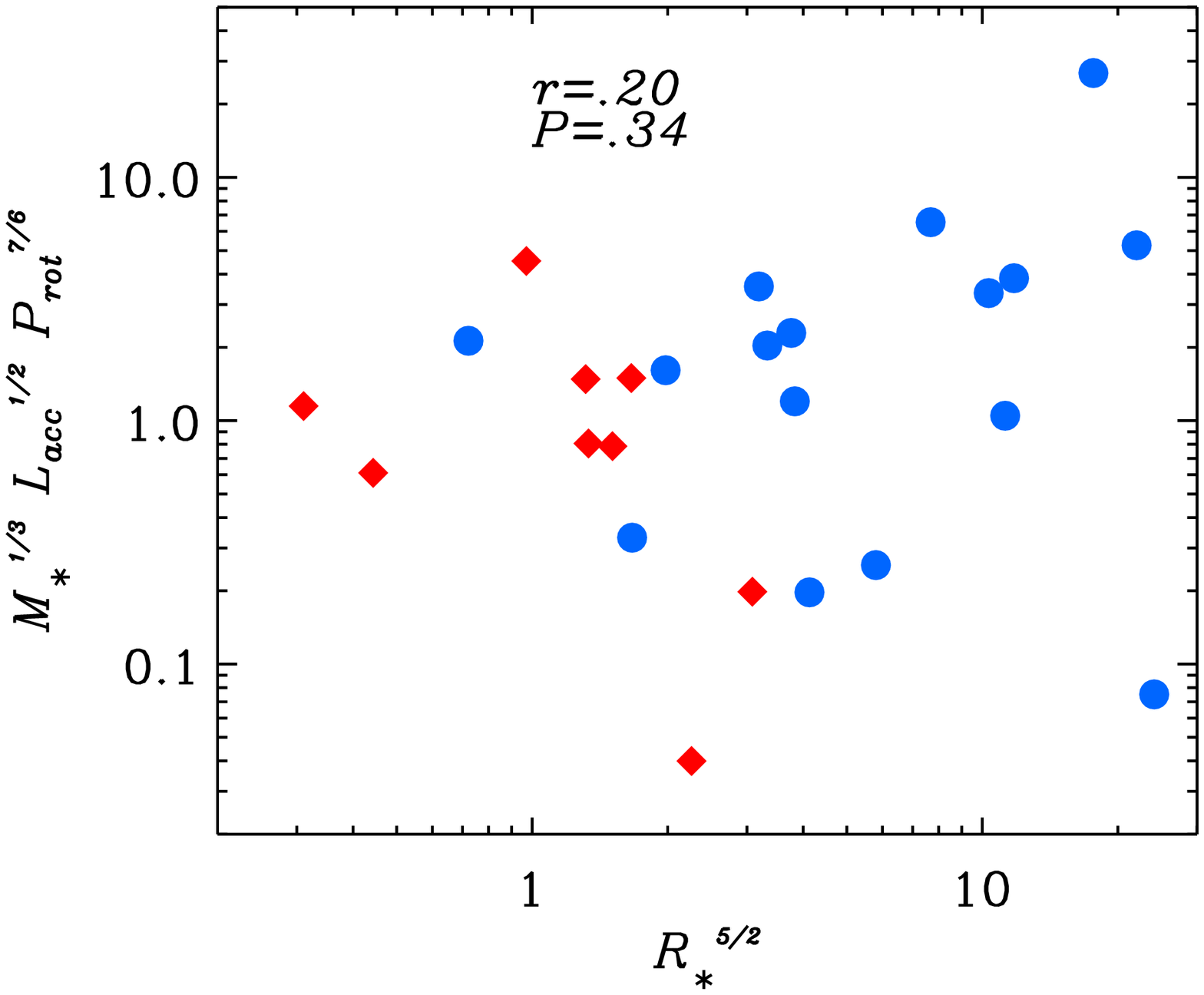}            
\caption{Equation 12 plotted for the SA (blue circles) and WA (red diamonds) stars.  No correlation is present, reaffirming the lack of support for the pure dipole models.}
\end{center}
\end{figure}
\clearpage
\begin{figure}[hbt]
\figurenum{12}
\begin{center}
\includegraphics[scale=.80]{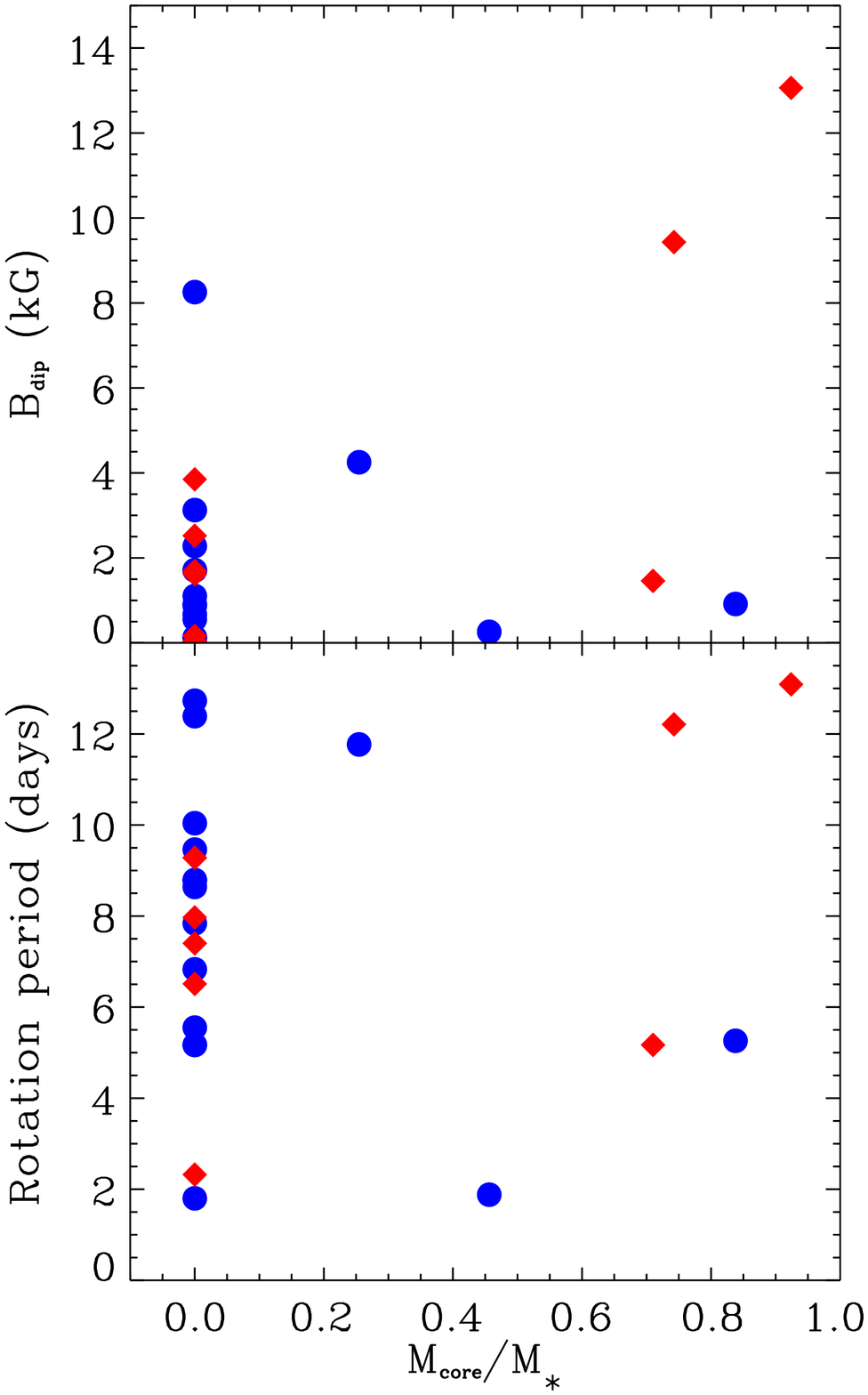}            
\caption{Theoretical equatorial dipolar magnetic field strengths at the stellar surface (upper panel) and rotation period (lower panel) plotted against the mass of the star's radiative core relative to the total stellar mass.  The field strengths are calculated using the relationship given in \citet{JK99} for the \citet{shu94} model, i.e. eq. (1) with the appropriate constants included.  Only the SA (blue circles) and WA stars (red diamonds) are included in the plot.  Kolmogorov--Smirnov tests for the fully convective vs. partially convective stars yield high probabilities ($P_{KS}$=.24, upper panel; $P_{KS}$=.38, lower panel) that the samples are drawn from the same distribution.}
\end{center}
\end{figure}
\clearpage
\begin{figure}[tbp]
\begin{center}
\figurenum{A1}
\includegraphics[scale=.90]{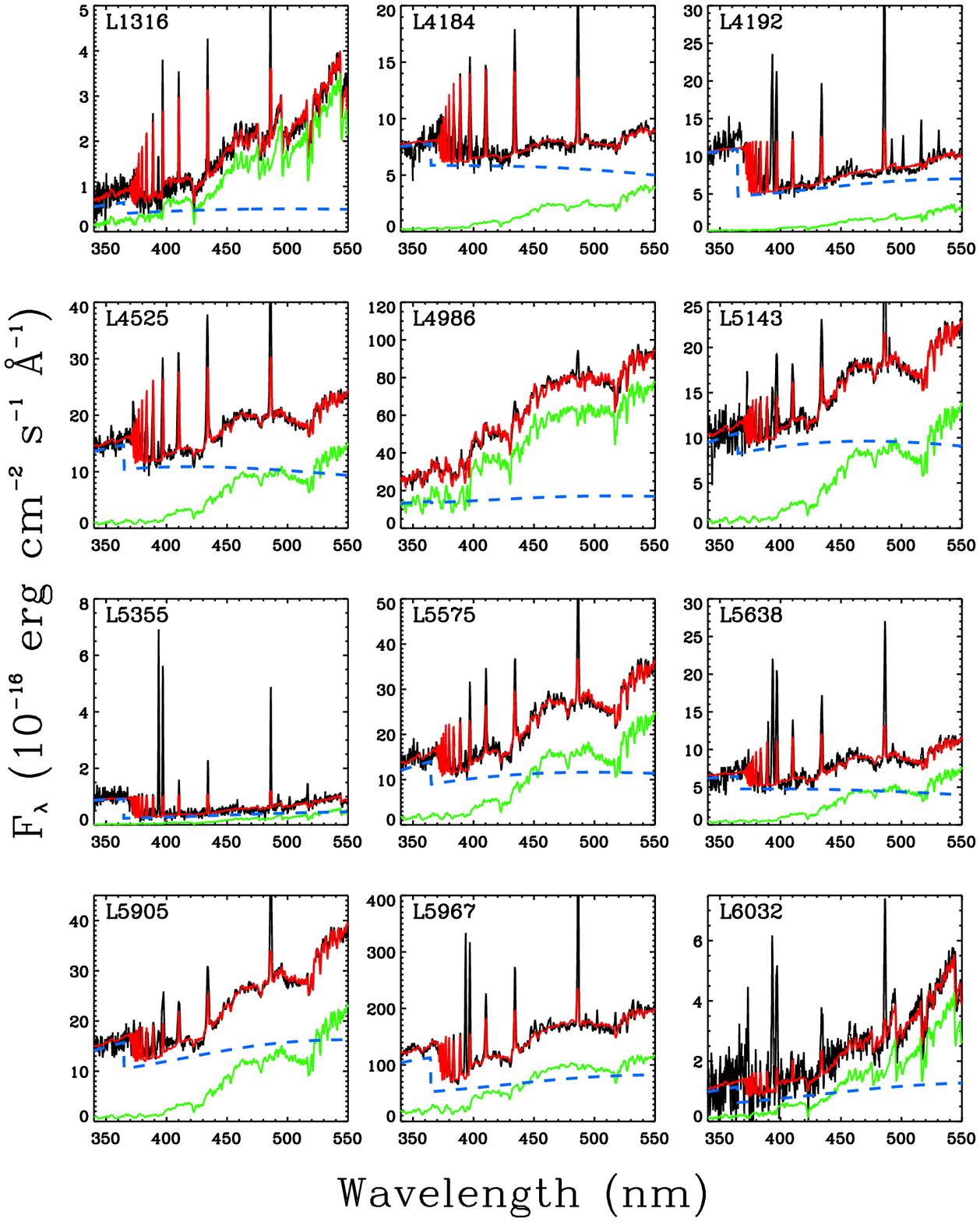}            
\caption{Flux--calibrated spectra and model fits for our entire NGC 2264 sample.  The SA stars are displayed first, followed by the WA and NA groups.  The colors and line--types are the same as in Fig. 1.}
\end{center}
\end{figure}
\clearpage
\begin{figure}[tbp]
\begin{center}
\figurenum{A1 cont}
\includegraphics[scale=.90]{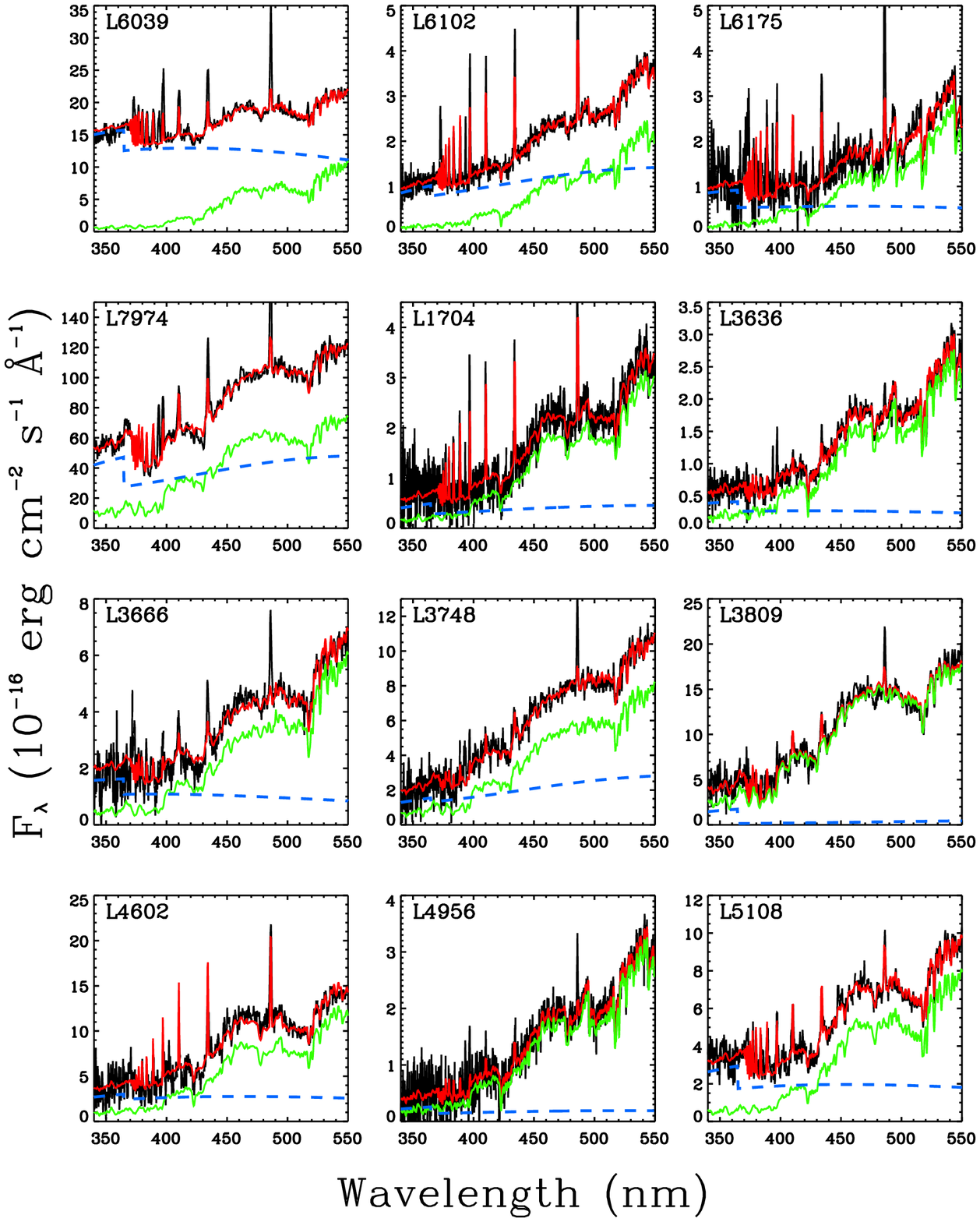}            
\end{center}
\end{figure}
\clearpage
\begin{figure}[tbp]
\begin{center}
\figurenum{A1 cont}
\includegraphics[scale=.90]{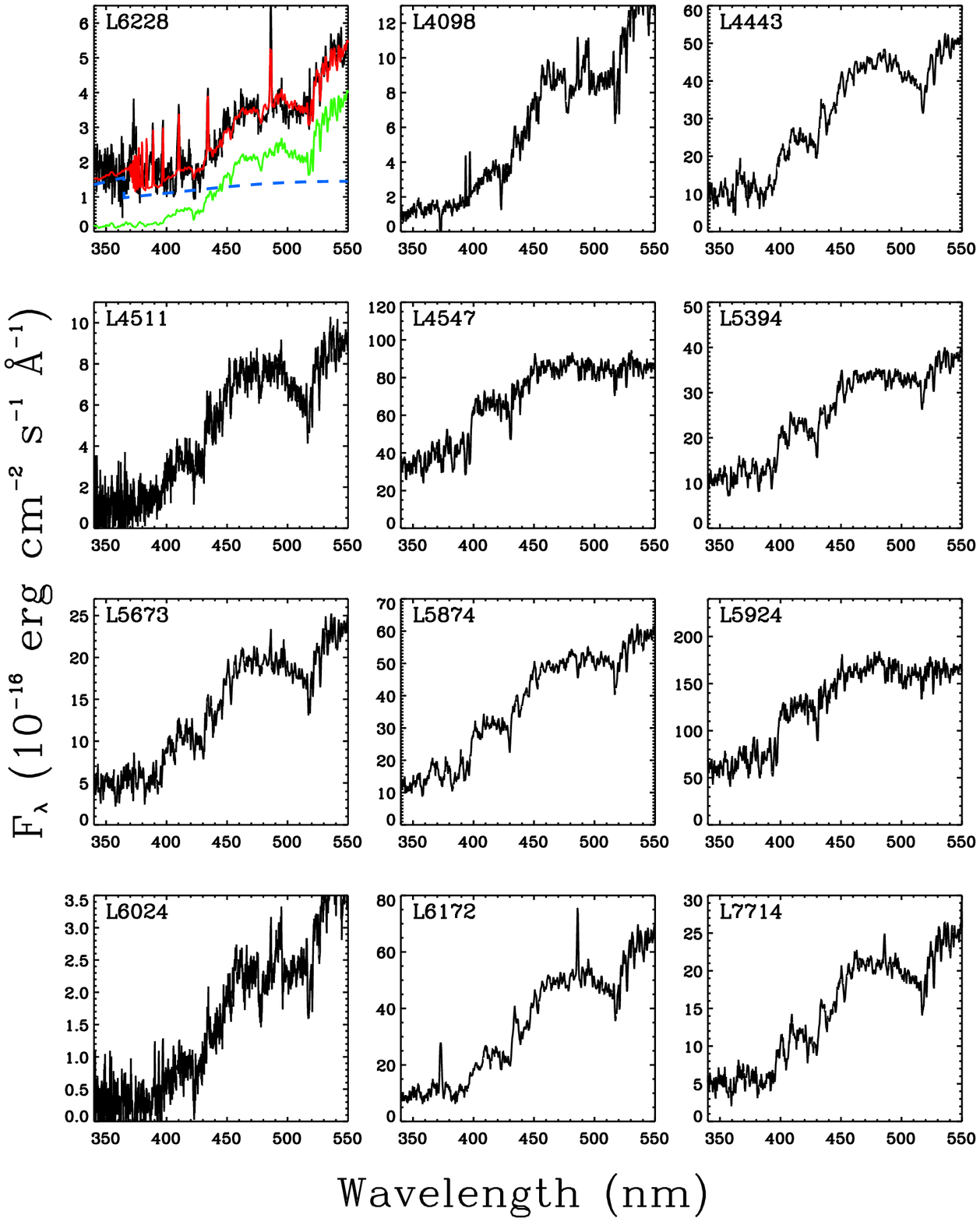}            
\end{center}
\end{figure}
\clearpage
\begin{figure}[tbp]
\begin{center}
\figurenum{B1}
\includegraphics[scale=.80]{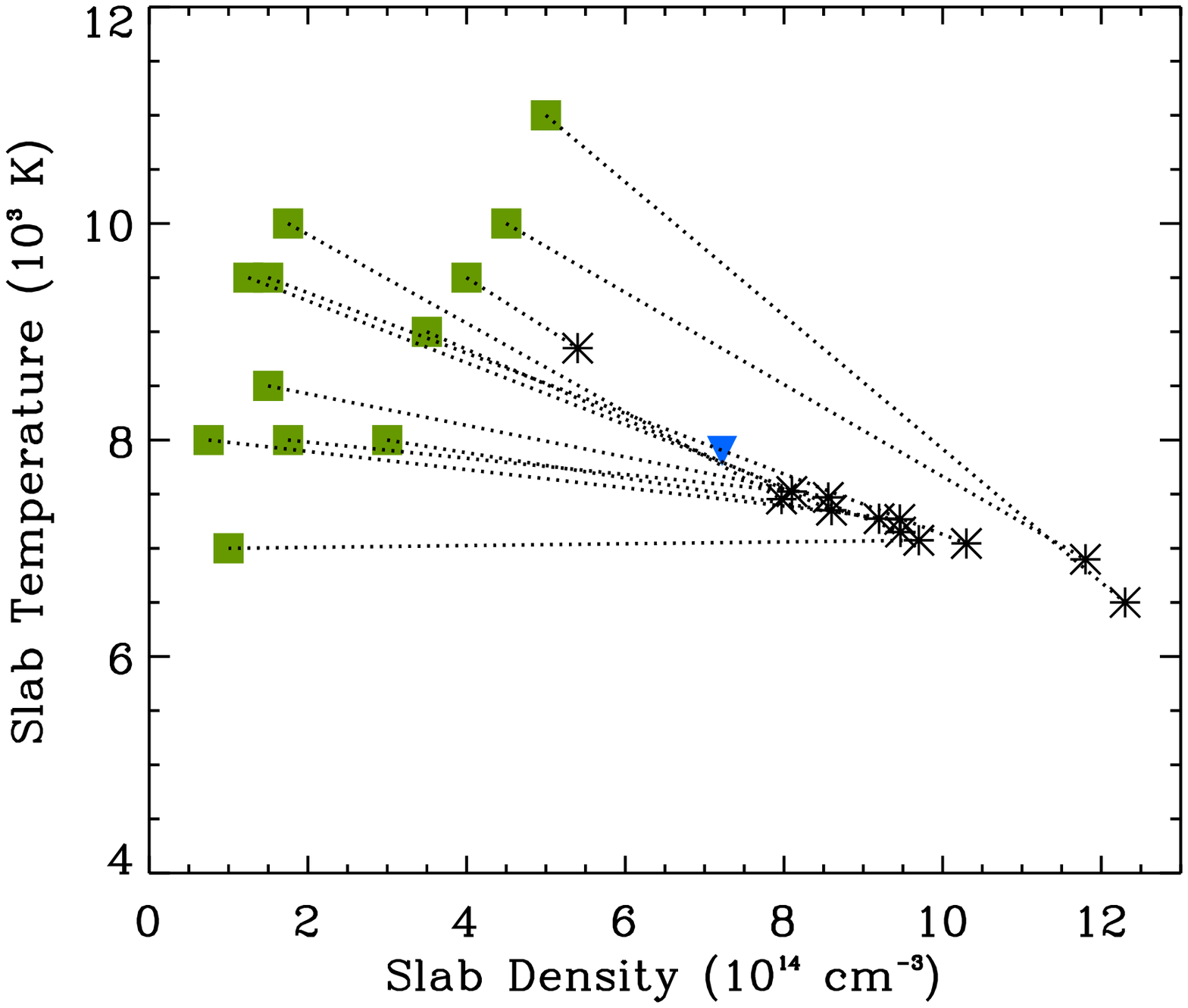}            
\caption{The effect of initial parameter choice on the final model values for L1316.  The initial values of $T$ and $n$ are represented by green squares; the final values are shown with black asterisks; the values used in the final model are plotted with a blue upsidedown triangle.  The dotted lines link the initial parameters with their final values.  The final values lie along a straight line indicating the need for another constraint on the model.}
\end{center}
\end{figure}
\clearpage
\begin{figure}[tbp]
\begin{center}
\figurenum{B2}
\includegraphics[scale=.80]{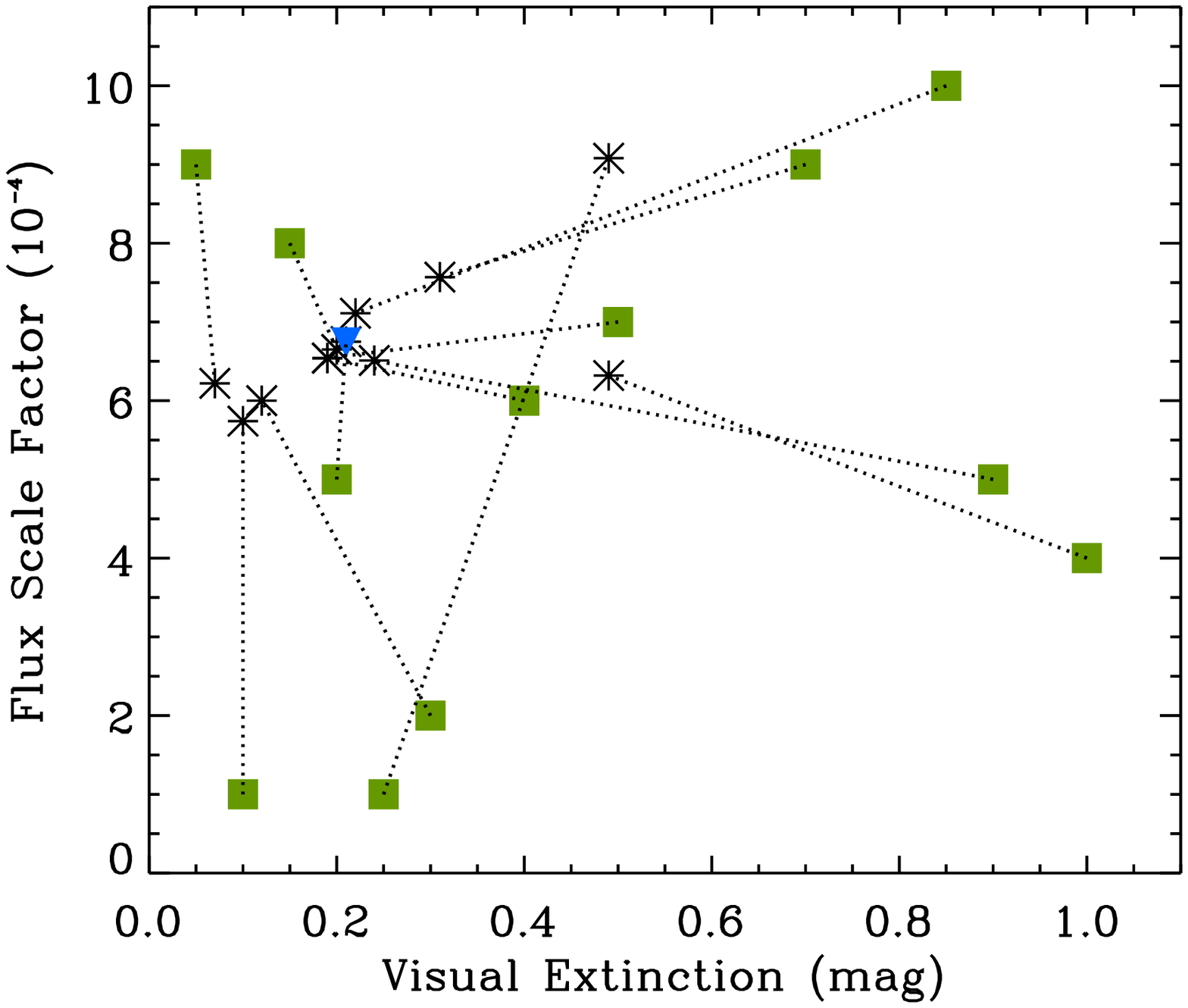}            
\caption{Same as Figure B1 but for \textit{A$_V$} and $\beta$.  Similar behavior is found for each pair of strongly coupled parameters.}
\end{center}
\end{figure}
\clearpage
\begin{deluxetable}{lcccl}
\tablecaption{Observing log}
\tablewidth{0pt}
\tablehead{\colhead{Observatory}&\colhead{Night}&\colhead{Conditions}&\colhead{Slit width}&\colhead{Instrument}}
\startdata
McDonald &28-Jan-2004&Photometric & 4\arcsec &LCS\\
HJS 2.7m &29-Jan-2004&Light clouds& ''       & ''\\
         &30-Jan-2004&Photometric & ''       & ''\\
\\
\hline
\\
Kitt Peak&26-Feb-2005&Clear       & 3\arcsec &RCS\\
Mayall 4m&27-Feb-2005&Closed      &          &   \\
         &28-Feb-2005&Mostly clear& 4\arcsec & ''\\
         &01-Mar-2005&Mostly clear& ''       & ''\\
\\
\hline
\enddata
\end{deluxetable}
          
\clearpage
\begin{deluxetable}{lcccc}
\tablecaption{NGC 2264 Sample}
\tablewidth{0pt}
\tablehead{\colhead{}&\multicolumn{2}{c}{Spectral Type}&\colhead{}&\colhead{}\\
\cline{2-3}\\
\colhead{ID}&\colhead{Literature$^{a}$}&\colhead{Best--fit$^{b}$}&\colhead{V--mag}&\colhead{Group$^c$}}
\tabletypesize{\footnotesize}
\startdata
L1316&M? &M3.5&17.55&SA\\
L4184&M2 &M0&16.67&SA \\
L4192&K4 &M2&16.28&SA \\
L4525&K6 &K7&15.54&SA\\
L4986&K1 &K1&14.14&SA\\
L5143&K4 &K7&15.37&SA\\
L5355&M5 &M3.5&19.12&SA \\
L5575&M1 &K7&15.09&SA \\
L5638&\nodata&K7&16.06&SA \\
L5905&K4 &K7&15.07&SA \\
L5967&\nodata&K3&15.37&SA \\
L6032&M3 &M3.5&17.53&SA\\
L6039&\nodata&K7&16.43&SA \\
L6102&K7 &M2&17.44&SA \\
L6175&M7 &M3.5&17.59&SA \\
L7974&K4 &K3&13.86&SA \\
L1704&M2 &M0&17.07&WA\\
L3636&M3 &M2&17.69&WA\\
L3666&K7 &K4&16.66&WA\\
L3748&K1 &K2&16.03&WA\\
L3809&K4 &K3&15.60&WA\\
L4602&K4 &M0&15.89&WA\\
L4956&M3 &M2&17.47&WA\\
L5108&K7 &K7&16.22&WA \\
L6228&M1 &K7&16.59&WA \\
L4098&M1.5&M2&16.04&NA\\
L4443&K4 &K3&14.58&NA\\
L4511&\nodata&K7&16.53&NA\\
L4547&G3 &K1&13.57&NA\\
L5394&K1 &K1&14.70&NA\\
L5673&K4 &K3&15.39&NA\\
L5874&K1 &K1&14.36&NA\\
L5924&G3 &K1&13.52&NA\\
L6024&\nodata&M2&17.82&NA\\
L6172&K5 &K3&14.47&NA\\
L7714&K4 &K2&15.44&NA\\
\hline
\enddata
\tablenotetext{a}{Spectral types determined by Rebull et al. (2002)}
\tablenotetext{b}{Spectral type of template star used to model the data.}
\tablenotetext{c}{SA=strongly accreting star, WA=weak accretor, NA=negligible accretor}
\end{deluxetable}
\clearpage
\begin{deluxetable}{lcccc}
\tablecaption{Main Sequence Templates}
\tablewidth{0pt}
\tabletypesize{\small}
\tablehead{\colhead{Name}&\colhead{Spectral Type}&\colhead{$B-V$}&\colhead{$R_T$}&\colhead{Distance}\\
\colhead{}&\colhead{}&\colhead{}&\colhead{(R$_\odot$)}&\colhead{(pc)}}
\startdata
HD10476   & K1 &0.84& .86 & 7.47\\
HD109011  & K2 &0.94& .83 &23.74\\
HD45088   & K3 &0.97& .82 &14.66\\
GL570A    & K4 &1.11& .79 & 5.91\\
GL394     & K7 &1.34& .67 &10.99\\
LTT11085  & M0 &\nodata& .41 &30.48\\
GJ393$^a$ & M2 &1.52& .51 & 7.23\\
GJ273$^a$ & M4 &1.57& .34 & 3.80\\
\hline
\enddata
\tablecomments{$R_T$ is the radius of the template star.}
\tablenotetext{a}{Observed at Kitt Peak}
\end{deluxetable}
\clearpage
\begin{deluxetable}{cccccccccc}
\tablecolumns{10}
\tablehead{\colhead{}&\multicolumn{4}{c}{K7 Main sequence template}&\colhead{}&\multicolumn{4}{c}{V830 Tau}\\
\cline{2-5} \cline{7-10}\\
\colhead{Star}&\colhead{$R_*$}&\colhead{$\dot{M}$}&\colhead{$f_{acc}$}&\colhead{$A_V$}&\colhead{}&\colhead{$R_*$}&\colhead{$\dot{M}$}&\colhead{$f_{acc}$}&\colhead{$A_V$}\\
\colhead{}&\colhead{(R$_\odot$)}&\colhead{(10$^{-8}$ M$_\odot$ yr$^{-1}$)}&\colhead{}&\colhead{(mags)}&\colhead{}&\colhead{(R$_\odot$)}&\colhead{(10$^{-8}$ M$_\odot$ yr$^{-1}$)}&\colhead{}&\colhead{(mags)}}
\tablewidth{0pt}
\tabletypesize{\footnotesize}
\tablecaption{Parameter Comparison}
\startdata
L4525&1.59&.87&.021&.05&&1.62&.94&.026&.42\\
L5108&1.12&.10&.012&$\sim$0&&1.21&.10&.006&.46\\
L5143&1.70&1.04&.024&.23&&1.66&.85&.023&.48\\
L5575&2.03&1.26&.032&.03&&2.11&1.35&.029&.45\\
L5638&1.31&.43&.010&.36&&1.12&.32&.038&.52\\
L5905&3.44&10.0&.018&1.08&&2.94&5.56&.028&1.18\\
L6039&1.62&1.42&.019&.32&&1.75&1.76&.017&.62\\
L6228&1.22&.22&.014&.93&&1.26&.23&.004&1.16\\
\hline
\enddata
\tablecomments{A comparison between the stellar and accretion parameters derived using a MS K7 template and those derived using V830 Tau, a WTTS template.}
\end{deluxetable}
\clearpage
\begin{deluxetable}{ccccccccccc}
\tablehead{\colhead{Star}&\colhead{$M_*$}&\colhead{$R_*$}&\colhead{$L_*$}&\colhead{$P_{rot}$}&\colhead{$A_V$}&\colhead{Age}&\colhead{$T_{eff}$}&\colhead{\textit{f$_{acc}$}}&\colhead{\textit{L$_{acc}$}}&\colhead{$\dot{M}$}\\
\colhead{}&\colhead{(M$_\odot$)}&\colhead{(R$_\odot$)}&\colhead{(L$_\odot$)}&\colhead{(days)}&\colhead{(mags)}&\colhead{(Myr)}&\colhead{(K)}&\colhead{}&\colhead{(L$_\odot$)}&\colhead{(10$^{-8}$ M$_\odot$ yr$^{-1}$)}}
\tablewidth{0pt}
\tabletypesize{\footnotesize}
\tablecaption{Stellar \& Accretion Parameters -- Strongly Accreting Stars}
\startdata
L1316&.16&1.76&.28&4.55&.21&1.0&3175&.003&.004&.20\\
L4184&.40&.88&.14&7.79&.61&7.0&3800&.028&.069&.73\\
L4192&.25&2.54&.78&7.79&1.33&0.5&3400&.009&.234&11.4\\
L4525&.56&1.59&.58&11.39&.05&2.0&4000&.021&.064&.87\\
L4986&1.85&2.68&4.36&4.26&1.18&3.0&5100&.020&.334&2.32\\
L5143&.54&1.70&.66&7.64&.23&3.0&4000&.024&.069&1.04\\
L5355&.12&1.23&.14&4.17&1.26&1.0&3175&.003&.016&.79\\
L5575&.52&2.26&1.17&11.73&.16&1.0&4000&.014&.073&1.26\\
L5638&.60&1.31&.40&6.84&.36&4.5&4000&.010&.041&.43\\
L5905&.50&3.44&2.72&8.46&1.08&0.4&4000&.018&.301&10.0\\
L5967&1.65&3.15&4.73&10.77&1.24&1.0&4800&.089&2.01&18.4\\
L6032&.18&3.57&1.16&.80&1.31&0.1&3175&.002&.030&2.83\\
L6039&.55&1.62&.60&5.83&.32&3.0&4000&.019&.101&1.42\\
L6102&.22&1.71&.35&9.04&1.08&1.5&3400&.012&.023&.87\\
L6175&.16&2.02&.37&4.17&.64&0.5&3175&.001&.008&.48\\
L7974&1.60&2.63&3.30&.88&1.33&2.0&4800&.099&1.08&8.53\\
\hline
\enddata
\end{deluxetable}
\clearpage
\begin{deluxetable}{ccccccccccc}
\tablehead{\colhead{Star}&\colhead{$M_*$}&\colhead{$R_*$}&\colhead{$L_*$}&\colhead{$P_{rot}$}&\colhead{$A_V$}&\colhead{Age}&\colhead{$T_{eff}$}&\colhead{\textit{f$_{acc}$}}&\colhead{\textit{L$_{acc}$}}&\colhead{$\dot{M}$}\\
\colhead{}&\colhead{(M$_\odot$)}&\colhead{(R$_\odot$)}&\colhead{(L$_\odot$)}&\colhead{(days)}&\colhead{(mags)}&\colhead{(Myr)}&\colhead{(K)}&\colhead{}&\colhead{(L$_\odot$)}&\colhead{(10$^{-8}$ M$_\odot$ yr$^{-1}$)}}
\tablewidth{0pt}
\tabletypesize{\footnotesize}
\tablecaption{Stellar \& Accretion Parameters--Weak Accretors}
\startdata
L1704&.40&.72&.10&8.28&.59&11.0&3800&.037&.005&.04\\
L3636&.21&1.33&.21&1.32&.41&2.0&3400&.001&.002&.06\\
L3666&.66&.60&.14&11.21&.01&25.0&4560&.009&.006&.02\\
L3748&.92&.99&.53&12.09&1.53&30.0&4960&.072&.065&.33\\
L3809&1.00&1.18&.66&4.17&1.30&11.0&4800&.018&.022&.12\\
L4602&.40&1.12&.23&6.97&.01&4.0&3800&.034&.044&.15\\
L4956&.21&1.57&.30&5.51&.49&2.0&3400&.001&.002&.06\\
L5108&.61&1.12&.29&6.40&$\sim$0&6.0&4000&.012&.012&.10\\
L6228&.60&1.22&.34&8.28&.93&5.0&4000&.014&.023&.22\\
\hline
\enddata
\end{deluxetable}
\clearpage
\begin{deluxetable}{cccccccc}
\tablehead{\colhead{Star}&\colhead{$M_*$}&\colhead{$R_*$}&\colhead{$L_*$}&\colhead{$A_V$}&\colhead{$P_{rot}$}&\colhead{Age}&\colhead{$T_{eff}$}\\
\colhead{}&\colhead{(M$_\odot$)}&\colhead{(R$_\odot$)}&\colhead{(L$_\odot$)}&\colhead{(mags)}&\colhead{(days)}&\colhead{(Myr)}&\colhead{(K)}}
\tablewidth{0pt}
\tabletypesize{\footnotesize}
\tablecaption{Stellar Parameters--Negligible Accretors}
\startdata
L4098&.24&1.35&.35&.32&2.71&.30&3400\\
L4443&1.30&1.32&.94&.27&4.50&5.0&4800\\
L4511&.60&1.06&.26&.01&12.47&7.0&4000\\
L4547&1.30&1.61&1.57&.06&3.42&8.0&5100\\
L5394&1.20&1.45&1.28&.64&2.38&10.0&5100\\
L5673&.90&.98&.53&.47&8.28&50.0&4960\\
L5874&1.55&2.05&2.55&.89&3.70&5.0&5100\\
L5924&1.70&2.31&3.24&.02&3.00&3.0&5100\\
L6024&.22&.80&.12&.56&9.71&1.0&3400\\
L6172&1.60&2.66&3.37&1.65&1.76&2.0&4800\\
L7714&1.10&1.30&.80&1.14&4.45&10.0&4800\\
\hline
\enddata
\end{deluxetable}
\clearpage
\begin{deluxetable}{lcc}
\tablehead{\colhead{Star}&\colhead{$F_{H\beta}$$^a$}&\colhead{$L_{H\beta}$}\\
\colhead{}&\colhead{(10$^{-14}$ erg cm$^{-2}$ s$^{-1}$)}&\colhead{(10$^{-5}$ L$_\odot$)}}
\tablewidth{0pt}
\tabletypesize{\footnotesize}
\tablecaption{WTTS H$\beta$ line luminosities}
\startdata
LkCa 3&2.42&1.48\\
V826 Tau&6.09&3.72\\
LkCa 4&5.86&3.57\\
V827 Tau&6.22&3.79\\
V830 Tau&4.08&2.49\\
\hline
\enddata
\tablecomments{H$\beta$ line luminosities for the WTTSs used to determine the threshold in Figure 5 (dotted line).  These stars show no evidence of any near--IR excess \citep{kenyon95} and only upper limits are found for the millimeter flux \citep{dutrey96}.  The luminosities were calculated assuming a distance of 140 pc to the Taurus star forming region.}
\tablenotetext{a}{\citet{valenti93}.}
\end{deluxetable}
\clearpage
\begin{deluxetable}{ccccccccccccc}
\tablecolumns{13}
\tablehead{\colhead{}&\multicolumn{5}{c}{R02$^{a}$}&\colhead{}&\multicolumn{5}{c}{This study}\\
\cline{2-6} \cline{8-12}\\
\colhead{Star}&\colhead{$M_*$}&\colhead{log(Age)}&\colhead{$R_*$}&\colhead{$A_V$}&\colhead{log($\dot{M}$)}&\colhead{}&\colhead{$M_*$}&\colhead{log(Age)}&\colhead{$R_*$}&\colhead{$A_V$}&\colhead{log($\dot{M}$)}&\colhead{Group}\\
\colhead{}&\colhead{(M$_\odot$)}&\colhead{(yrs)}&\colhead{($R_\odot$)}&\colhead{(mags)}&\colhead{(M$_\odot$ yr$^{-1}$)}&\colhead{}&\colhead{(M$_\odot$)}&\colhead{(yrs)}&\colhead{($R_\odot$)}&\colhead{(mags)}&\colhead{(M$_\odot$ yr$^{-1}$)}&\colhead{}}
\tablewidth{0pt}
\tabletypesize{\footnotesize}
\tablecaption{Parameter Comparison}
\rotate
\startdata
L4184&.29&5.6&2.03&.67&-7.99&&.40&6.8&.88&.61&-8.14&SA\\
L4525&.68&6.0&1.93&\nodata&-8.24&&.56&6.3&1.59&.05&-8.06&SA\\
L4986&1.46&6.0&2.66&1.03&-7.57&&1.85&6.5&2.68&1.18&-7.63&SA\\
L5143&.65&6.0&1.90&.61&-7.72&&.54&6.5&1.70&.23&-7.98&SA\\
L1704&.35&6.0&1.56&.43&-8.73&&.40&7.0&.72&.59&-9.37&WA\\
L3636&.32&6.2&1.30&.28&-8.79&&.21&6.3&1.33&.41&-9.21&WA\\
L3748&.74&5.7&2.43&\nodata&-8.25&&.92&7.5&.99&1.53&-8.48&WA\\
L3809&.81&6.4&1.50&.36&-8.31&&1.00&7.0&1.18&1.30&-8.90&WA\\
L4602&.65&6.2&1.56&.44&\nodata&&.40&6.6&1.12&.01&-8.81&WA\\
L4956&.31&6.0&1.49&.26&-8.48&&.21&6.3&1.57&.49&-9.24&WA\\
L5108&.66&6.1&1.71&.98&-8.61&&.61&6.8&1.22&$\sim$0&-8.98&WA\\
L4098&.27&5.2&2.40&.34&\nodata&&.24&5.5&2.58&.32&\nodata&NA\\
L4547&1.81&6.7&2.13&1.03&\nodata&&1.30&6.9&1.61&.06&\nodata&NA\\
L5673&1.21&6.9&1.38&1.08&-8.88&&.90&7.7&.98&.47&\nodata&NA\\
L5874&1.40&6.4&1.95&.78&\nodata&&1.55&6.7&2.05&.89&\nodata&NA\\
L6172&.91&5.5&3.21&.57&-6.89&&1.60&6.3&2.66&1.65&\nodata&NA\\
\hline
\enddata
\tablecomments{A comparison between the stellar and accretion parameters derived by Rebull et al. (2002) and this study.  Most mass measurements agree to within 30\%.  R02 calculates much younger ages for our sample.  We attribute this to their use of the D'Antona \& Mazzitelli (1994) stellar evolotion models which they show (R02 Fig. 13) predict ages much younger than those of Siess et al. (2000).}
\tablenotetext{a}{Rebull et al. (2002)}
\end{deluxetable}
\clearpage
\begin{deluxetable}{lcccccc}
\tablecolumns{7}
\tablehead{\colhead{}&\multicolumn{2}{c}{Eq. 1}&\colhead{}&\multicolumn{3}{c}{Eq. 3}\\
\cline{2-3} \cline{5-7}\\
\colhead{}&\colhead{\textit{r}}&\colhead{$P$}&\colhead{}&\colhead{\textit{r}}&\colhead{$P$}&\colhead{Slope}}
\tablewidth{0pt}
\tabletypesize{\small}
\tablecaption{Summary of the correlation results}
\startdata
WA stars&-.30&.43&&.88&.002&.57$\pm$.06\\
SA stars&.26&.32&&.91&9x10$^{-7}$&.58$\pm$.06\\
All stars&.31&.13&&.91&3x10$^{-10}$&.63$\pm$.03\\
\hline
\enddata
\tablecomments{Summary of the correlation results for the assumption of a constant magnetic field from star to star.  Best--fit slopes for the poor correlations from eq. (1) are not computed.  The slope of the best--fit line to the data plotted using eq. (3) is significantly different than the predicted slope of 1.  \textit{r} is the linear correlation coefficient; $P$ is the correlation significance associated with the given value of \textit{r}.}
\end{deluxetable}
\clearpage
\begin{deluxetable}{lccc}
\tablecolumns{4}
\tablehead{\colhead{}&\colhead{\textit{r}}&\colhead{}&\colhead{$P$}}
\tablewidth{0pt}
\tabletypesize{\small}
\tablecaption{Correlation results for $\dot{M}$, $R_*$, and $f_{acc}$.}
\startdata
$\dot{M}$ vs. $R_*^2$&.81&&8.2x10$^{-7}$\\
$\dot{M}$ vs. $f_{acc}$&.41&&.044\\
$\dot{M}$ vs. $f_{acc}$$R_*^2$&.80&&1.3x10$^{-6}$\\
\hline
\enddata
\tablecomments{r is the linear correlation coefficient; $P$ is correlation significance associated with the given value of r.}
\end{deluxetable}
\clearpage
\begin{deluxetable}{ccccccccc}
\tablecaption{Parameter tests for L1316: $n$ vs. $T$}
\tablenum{B1}
\tablewidth{0pt}
\tabletypesize{\small}
\tablehead{\colhead{Trial}&\colhead{$n_0$}&\colhead{$T_0$}&\colhead{$n$}&\colhead{$T$}&\colhead{$\tau_{4000}$}&\colhead{$R_*$}&\colhead{$\dot{M}$}&\colhead{$\chi^2$}\\
\colhead{}&\colhead{(10$^{14}$ cm$^{-3}$)}&\colhead{(K)}&\colhead{(10$^{14}$ cm$^{-3}$)}&\colhead{(K)}&\colhead{}&\colhead{(R$_\odot$)}&\colhead{($10^{-8}$ M$_\odot$ $yr^{-1}$)}&\colhead{}}
\startdata
1&1.25&9500  &9.47&7150&.21&1.82&3.12&1.31\\
2&1.50&8500  &8.56&7470&.21&1.84&3.15&1.32\\
3&1.75&10000&7.97&7455&.21&1.84&3.13&1.32\\
4& .75&8000  &9.46&7267&.21&1.83&3.29&1.31\\
5&5.00&11000&12.3&6500&.22&1.75&3.06&1.30\\
6&3.00&8000&9.20&7273&.21&1.83&3.22&1.31\\
7&3.50&9000&8.60&7352&.21&1.82&3.13&1.31\\
8&4.00&9500&5.40&8849&.20&1.88&2.87&1.34\\
9&4.5&10000&11.80&6898&.21&1.80&3.21&1.31\\
10&1.00&7000&9.70&7074&.21&1.81&3.15&1.31\\
11&1.50&9500&10.30&7045&.21&1.81&3.27&1.31\\
12&1.75&8000&8.10&7524&.21&1.85&3.23&1.32\\
\hline
\enddata
\tablecomments{Initial and final parameter values for the density, $n$, and temperature, $T$.  Each trial corresponds to a pair of points in Fig. B1.}
\end{deluxetable}
\clearpage
\begin{deluxetable}{ccccccccc}
\tablenum{B2}
\tablecaption{Parameter tests for L1316: $A_V$ vs. $\beta$}
\tablewidth{0pt}
\tabletypesize{\small}
\tablehead{\colhead{Trial}&\colhead{$A_{V0}$}&\colhead{$\beta_0$}&\colhead{$A_V$}&\colhead{$\beta$}&\colhead{$\tau_{4000}$}&\colhead{$R_*$}&\colhead{$\dot{M}$}&\colhead{$\chi^2$}\\
\colhead{}&\colhead{(mags)}&\colhead{(10$^{-20}$}&\colhead{(mags)}&\colhead{(10$^{-20}$)}&\colhead{}&\colhead{(R$_\odot$)}&\colhead{($10^{-8}$ M$_\odot$ yr$^{-1}$)}&\colhead{}}
\startdata
1&.15&8.0&.20&6.65&.21&1.84&3.13&1.31\\
2&.20&5.0&.21&6.75&.20&1.86&3.18&1.32\\
3&.25&1.0&.49&9.08&.36&2.16&5.94&1.53\\
4&.30&2.0&.12&6.00&.22&1.75&3.19&1.30\\
5&.40&6.0&.19&6.54&.21&1.83&3.11&1.31\\
6&.50&7.0&.19&6.54&.21&1.83&3.12&1.31\\
7&.70&9.0&.31&7.57&.19&1.97&3.76&1.34\\
8&.85&10.0&.22&7.11&.27&1.91&3.25&1.45\\
9&.90&5.0&.24&6.51&.17&1.82&4.08&1.36\\
10&1.00&4.0&.49&6.32&.18&1.80&9.91&1.87\\
11&.05&9.0&.07&6.22&.24&1.78&2.36&1.40\\
12&.10&1.0&.10&5.74&.21&1.71&3.21&1.33\\
\hline
\enddata
\tablecomments{Same as Table B1 but for the flux scale factor $\beta$ and the visual extinction $A_V$.  These data correspond to Fig. B2.}
\end{deluxetable}

\end{document}